\documentclass[aps,prl,twocolumn,floatfix]{revtex4-2}
    
    \usepackage{graphicx}
    \usepackage{dcolumn}
    \usepackage{bm}
    \usepackage{amsmath}
    
    \usepackage{upgreek}
    \usepackage{empheq}
    
    \usepackage{soul}
    \usepackage{mathrsfs}
    \usepackage{romannum}
    \usepackage{xcolor}
    \usepackage{caption}

    \usepackage{caption}
    \usepackage{ragged2e}
    \usepackage{caption}
    \usepackage{xr}
\begin{document}

    \title{Spreading the Damage: How Heterogeneity Delocalizes Fracture in Polymer Networks}
    
    
    \author{Saleh Assadi}
    \author{Zachary White}
    \thanks{S.A. and Z.T.W. contributed equally to this work.}
    \author{Radhika Joshi}
    \author{Franck J. Vernerey}
    \email[Correspondence to: ]{franck.vernerey@colorado.edu}
    
    \affiliation{University of Colorado at Boulder, Boulder, Colorado, 80309, USA}
    
    
    \date{\today}
    
    \begin{abstract}
    Fracture in polymer networks is strongly influenced by how network architecture and topology organize damage in space. Although these materials can sustain large deformations under homogeneous loading, they often fail in a brittle and flaw-sensitive manner once defects are present. Here we show that network heterogeneity is one key architectural parameter controlling this transition. Using a discrete mesoscale network model that explicitly resolves strand-level prestretch, force transmission, and irreversible bond scission, we demonstrate that heterogeneity in prestretch and strand lengths broadens crack-induced deformation, delocalizes bond rupture, and enlarges the fracture process zone. In polydisperse networks, increasing contour-length disorder broadens the prestretch distribution and systematically increases both toughness and the intrinsic length scale governing flaw sensitivity. Bimodal networks amplify this mechanism further through combined contrasts in prestretch and strand lengths, producing the largest damage zones and the highest toughness among the architectures studied. Inspired by the results, we derive a scaling estimate relating the emergent fracture-active length to the spatial range of crack-induced deformation and the logarithmic spread of strands' remaining stretch to rupture.
    \end{abstract}
    
    \maketitle
    
    Soft polymer networks such as elastomers and hydrogels combine large deformability with permanent elasticity, enabling their widespread use in soft structural, biomedical, and bioinspired materials \cite{Creton2017,Lee_tissue_hydrogel,Biomed_implant_Tineke,MARS2002949}. Yet their fracture response is often governed by flaw sensitivity, whereby a small crack can trigger catastrophic failure at strains far below those achieved in defect-free loading. Classical molecular theories, most notably the Lake--Thomas model \cite{LakeThomas1967}, attribute fracture to the scission energy of load-bearing strands crossing the crack plane. In controlled tetra-PEG gels \cite{Sakai_scaling,Sakai_Akagi}, this local picture was validated quantitatively with a nearly constant prefactor, \(k\approx3\). More recent theories, however, account for nonlinear chain extension and energy stored in neighboring unbroken strands \cite{Wang2019LakeThomas,Wang2023UnbrokenStrands}, while simulations and loop-opening models reveal nonlocal elastic energy release extending beyond the crack plane \cite{Polymerlike_network_fracture,loop_opening_zhao,Network_fracture,Hartquist2026Fractocohesive}. Thus, the Lake--Thomas picture provides an important localized limit but does not explain why many networks fracture through zones extending far beyond a molecular layer.

    Extended fracture zones are well established in tough polymer networks. In double- and multiple-network hydrogels and elastomers, sacrificial bond rupture and stress redistribution activate damage over a broad region surrounding the crack tip \cite{slootman2020quantifying,slootman_pnas,Gong_DN_2010,Gong_2003_DN_Extreme_strength}. Entanglements, interpenetrating networks, and other architectural features can similarly enhance flaw tolerance by redistributing deformation and scission \cite{Junsoo_entangle_science,Creton2017,Dookhith2025,creton_2021,Zhigang_perfect_vs_imperfect}. More broadly, network heterogeneity may arise through topology, spatial cross-link density, strand length and conformation, or local rupture strength. Coarse-grained and lattice models show that these forms of disorder reorganize load sharing and rupture statistics \cite{Lin2020FractureTopologicalDefects,Arora2020TopologicalDefects,Arora2022TopologicalInhomogeneities,Arora2025SpatialHeterogeneity,Beech2023Depercolation}. In particular, networks containing mixed strong and weak strands can exhibit nonmonotonic fracture energy, with certain mixtures outperforming both weak and strong networks \cite{Hartquist2025HybridStrength}.

    One important but distinct form of this broader disorder is equilibrium prestretch heterogeneity which arises from unequal strand polymerization and irregular network architecture \cite{ITSKOV2016512,Macromolecules_polydisperse_elasticity,Zhang_sanoja_MD_polydisp_fracture,Arora2026BimodalFracture,Arujo_Brassart_discrete}. Highly and weakly stretched strands coexist in the equilibrated reference state, even with the same force--extension law and intrinsic scission threshold. Their different remaining stretches to rupture make prestretch heterogeneity a state-based analogue of mixed bond strengths that changes when and where scission occurs.

    \begin{figure}[t!]
    \centering
    \includegraphics[width=1\linewidth, trim={1.1cm 1.0cm 1.0cm 0cm}]{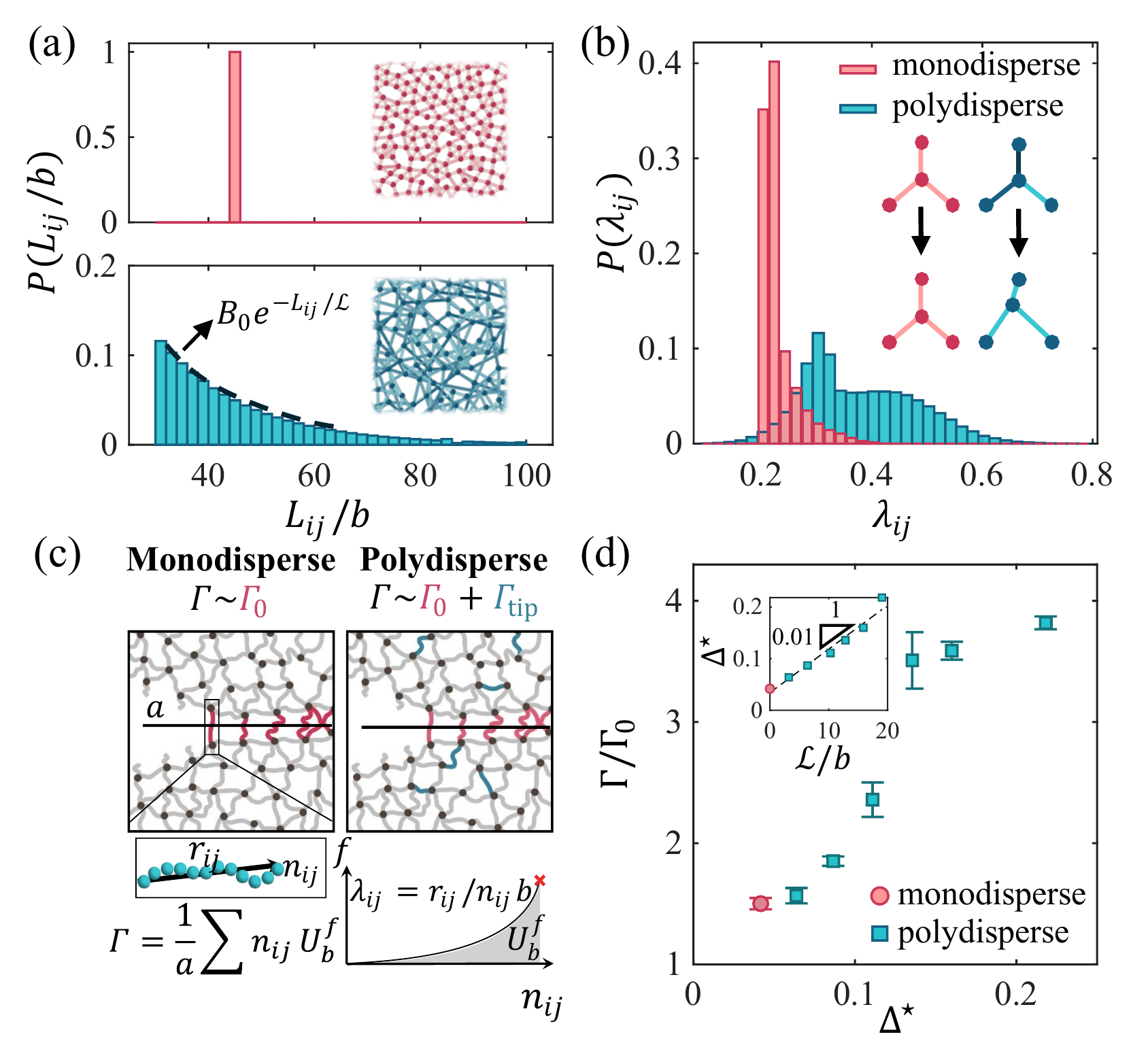}
    \caption{\footnotesize \justifying \textbf{Heterogeneity and toughness.} (a) Contour-length distributions \(P(L_{ij}/b)\) and representative equilibrated snapshots. (b) Equilibrium prestretch distributions \(P(\lambda_{ij})\), where \(\lambda_{ij}=r_{ij}/L_{ij}\) is the stretch of chain $i-j$. The schematics illustrate the heterogeneity upon equilibration. (c) Localized (monodisperse) and distributed (polydisperse) scission. (d) Normalized fracture toughness \(\Gamma/\Gamma_0\) versus the dimensionless prestretch width \(\Delta^\star\). The inset shows the relation between \(\Delta^\star\) and the contour-length disorder \(\mathcal{L}/b\).}
    \label{fig:introductory}
\end{figure}

    In this Letter, we investigate how network architecture, specifically prestretch and bond length distributions, control damage delocalization and fracture resistance. We employ a discrete mesoscale network model \cite{DN_toughness_PRL,Polymerlike_network_fracture,Network_fracture,Tauber_PRM_DN_failure,White2025,Arujo_Brassart_discrete,Hartquist2025HybridStrength,Assadi_mesoscale} that resolves strand-level force transmission and irreversible bond scission while accessing the length scales relevant to crack propagation. This approach retains rupture statistics homogenized in continuum damage and phase-field models \cite{LiBouklas2019,TehraniSarvestani2017,ZHAO2012319,Lamont_continuum_damage,ZHANG20151}, while reaching scales difficult for fully atomistic simulations \cite{Tauber_DN_MD,DN_MD_Higuchi,Gado_MD_pnas,Zhang_sanoja_MD_polydisp_fracture}. We show how equilibrium prestretch heterogeneity and the distances spanned by network strands jointly influence crack-tip deformation and distributed rupture, and develop a scaling estimate for the emergent fracture-active length.
    
   The elastomer is represented as a network of crosslinking nodes connected by elastic strands. A strand joining nodes \(i\) and \(j\) has end-to-end distance \(r_{ij}=\|\bm r_j-\bm r_i\|\) and contour length \(L_{ij}=n_{ij}b\), where \(n_{ij}\) is its number of Kuhn segments and \(b\) is the Kuhn length. Treating \(r_{ij}\) and \(n_{ij}\) as independent variables during network generation provides direct control over the initial strand stretch, \(\lambda_{ij}=r_{ij}/L_{ij}\), and permits the construction of monodisperse, polydisperse, and bimodal architectures at fixed network density. Strand elasticity follows the finitely extensible Cohen model \cite{Cohen1991}, and strands rupture irreversibly at \(\lambda_c=0.95\). Short-range repulsive and cohesive interactions \cite{Sanyal16,Sanyal18} provide excluded volume, near-incompressibility, and crack-face traction. Simulations are performed in LAMMPS \cite{THOMPSON2022108171} under quasistatic uniaxial loading following overdamped equilibration to nodal force balance. Full interaction laws and numerical procedures are given in the Supplemental Material.

    To systematically study the effect of contour-length heterogeneity, we construct two-dimensional monodisperse and polydisperse networks at fixed Kuhn segment density, using five independent realizations for each architecture. As illustrated in normalized contour length histograms [Fig. \ref{fig:introductory}(a)], monodisperse networks contain a single contour length value, whereas the polydisperse distributions follow \(P(L_{ij})\propto\exp(-L_{ij}/\mathcal{L})\) over the prescribed contour-length range \cite{ITSKOV2016512,MULDERRIG2021103857}. The generation procedure varies the breadth through \(\mathcal{L}\) while keeping the mean contour length nearly unchanged.

    Equilibration to minimum energy, converts this contour-length disorder into prestretch heterogeneity. Monodisperse networks retain a narrow prestretch distribution \(P(\lambda_{ij})\), whereas polydisperse networks develop a broad distribution in which highly stretched short strands coexist with weakly stretched long strands [Fig.~\ref{fig:introductory}(b)]. We quantify this breadth by the dimensionless measure \(\Delta^\star=\sigma_\lambda/(\lambda_c-\langle\lambda\rangle)\), where \(\sigma_\lambda\) and \(\langle\lambda\rangle\) are the standard deviation and mean of the equilibrated prestretch distribution, respectively. The inset of Fig.~\ref{fig:introductory}(d) shows that \(\Delta^\star\) increases approximately linearly with \(\mathcal{L}/b\), directly connecting the imposed contour-length disorder to the prestretch heterogeneity.

    We next perform mode-I fracture simulations on single-edge notched samples. The stress stretch curves for these simulations are provided in the Supplemental Material. Then, the fracture toughness is computed from the scission energy of all failed strands per unit crack advance, \(\Gamma=\frac{1}{a}\sum n_{ij} U_b^f\), as illustrated in Fig.~\ref{fig:introductory}(c). Here, \(n_{ij}\) is the number of Kuhn segments in failed strand \(i-j\), \(U_b^f\) is the rupture energy per Kuhn segment, and \(a\) is the advanced crack area. We normalize this quantity by the Lake--Thomas fracture toughness, \(\Gamma_0\sim\nu\langle n\rangle U_b^f\xi\), where \(\nu\) is the crosslink density, $\langle n\rangle$ is the mean number of Kuhn segments per chain and \(\xi\) is the average mesh size \cite{LakeThomas1967,Sakai_scaling,Sakai_Akagi}. The average mesh size \(\xi\) is measured as the mean nearest-neighbor intercrosslinker distance  of the equilibrated reference network. We also define the mean strand end-to-end distance, \(\langle\ell\rangle=\langle r_{ij}^{\mathrm{ref}}\rangle\), where the average includes all intact bonded strands. Thus, \(\xi\) characterizes local crosslink spacing which is nearly constant across architectures due to fixed density, whereas \(\langle\ell\rangle\) measures the mean distance spanned by bonded connections.

    Figure~\ref{fig:introductory}(d) shows that normalized fracture toughness \(\Gamma/\Gamma_0\) increases systematically with \(\Delta^\star\) over the range studied, from approximately \(1.5\) in the monodisperse network to nearly \(4\) in the most heterogeneous polydisperse networks. Figure~\ref{fig:introductory}(c) schematic suggests the physical origin of this enhancement. In the monodisperse network, scission remains concentrated along the advancing crack plane, giving a predominantly localized contribution, consistent with Lake-Thomas local estimate \(\Gamma_0\). In polydisperse networks, highly prestretched strands require less additional deformation to reach \(\lambda_c\) and can therefore rupture away from the nominal crack path \cite{Long2020TheFO}. This distributed scission produces the additional contribution denoted by \(\Gamma_{\mathrm{tip}}\). Figure~\ref{fig:introductory} establishes this correlation but the underlying spatial mechanism and scaling are examined next.

\begin{figure}[t!]
    \centering
    \includegraphics[width=1.0\linewidth, trim={1.1cm 0.8cm 1.0cm 0cm}]{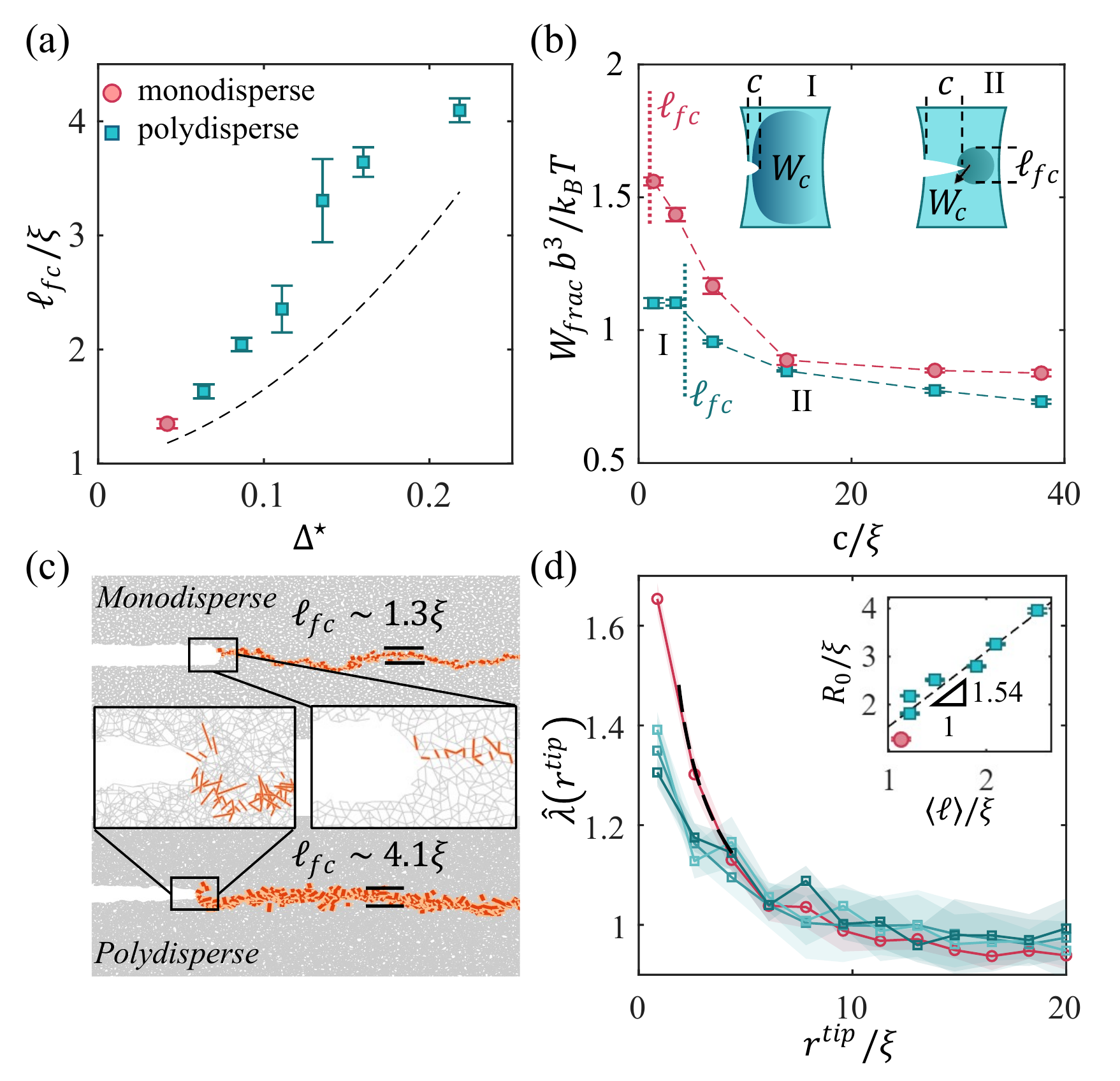}
    \caption{\footnotesize \justifying \textbf{Fracture length and flaw tolerance.} (a) Normalized fractocohesive length \(\ell_{fc}/\xi\) versus prestretch heterogeneity \(\Delta^\star\). The dashed curve is the prediction of the scaling analysis. (b) Normalized work of fracture \(W_{\mathrm{frac}}b^3/k_BT\) versus normalized notch length \(c/\xi\). Vertical lines mark \(\ell_{fc}\) and separate the flaw-insensitive (I) and flaw-sensitive (II) regimes. (c) Reference-configuration damage maps for representative monodisperse and polydisperse networks. (d) Normalized strand-stretch profiles \(\widehat{\lambda}(r^{\mathrm{tip}}) = \lambda(r^{tip})/\langle\lambda\rangle \) ahead of the crack tip for increasing \(\Delta^\star\). The inset relates \(R_0/\xi\) to \(\langle\ell\rangle/\xi\), with a linear slope of approximately \(1.54\).}
    \label{fig:fractocohesive_length}
\end{figure}

    To determine whether the larger fracture toughness corresponds to a larger fracture-active region, we measure the fractocohesive length \(\ell_{fc}=\Gamma/W_c\) \cite{Long2020TheFO,Zhigang_chem_review}, where \(W_c\) is the critical work density obtained by integrating the unnotched stress--stretch curve up to fracture. It increases from approximately \(1.3\xi\) in monodisperse networks to \(4.1\xi\) in the most heterogeneous polydisperse networks [Fig.~\ref{fig:fractocohesive_length}(a)]. Damage maps independently show a corresponding transition from localized to distributed scission [Fig.~\ref{fig:fractocohesive_length}(c)].

    The origin of this enlarged damage zone is resolved by measuring the stretch profiles as a function radial distance from crack tip $\lambda (r^{tip})$ in Fig.~\ref{fig:fractocohesive_length}(d). The curves from top to bottom respectively correspond to $\Delta^\star=0.04$ (monodisperse), $\Delta^\star=0.08$, $\Delta^\star=0.16$ and $\Delta^\star=0.21$. At common applied global stretch of $1.5$, increasing \(\Delta^\star\) produces a progressively slower normalized decay away from the crack tip. The profiles are fitted by \(A\exp(-r^{\mathrm{tip}}/R_0)+A_\infty\), where \(R_0\) is the crack-tip stretch-decay lengthscale. The inset of Fig.~\ref{fig:fractocohesive_length}(d) shows an approximately proportional relation between stretch-decay lengthscale \(R_0/\xi\) and mean strand end to end distance \(\langle\ell\rangle/\xi\), motivating the estimate \(R_0\simeq c_\ell\langle\ell\rangle\), with \(c_\ell=1.54\). This relation connects the spatial range of crack-induced deformation to the mean distance spanned by bonded strands in the reference network.

    These observations provide the physical ingredients for a scaling estimate of damage broadening. A strand with equilibrium prestretch \(\lambda_{ij}\) has rupture reserve \(\rho_{ij}=\lambda_c-\lambda_{ij}>0\). Using the measured decay length \(R_0\) as an estimate of the transverse range, we approximate the additional stretch at reference distance \(y\) from the crack plane by \(\delta\lambda_{\mathrm{tip}}(y)\simeq\delta\lambda_0\exp(-y/R_0)\). Equating this increment to the reserve gives the activation coordinate \(Y_{ij}=R_0\ln(\delta\lambda_0/\rho_{ij})\), showing why highly prestretched strands with small reserves can rupture farther from the crack path. The common amplitude shifts every coordinate equally, giving \(\operatorname{std}_{\mathrm{all}}(Y)=R_0\Delta_{\log}\), where \(\Delta_{\log}=\langle\left(\ln\rho-\langle\ln\rho\rangle\right)^2\rangle^{1/2}\) measures the logarithmic reserve spread over all intact reference strands. The operator  $\langle .\rangle$ denotes the mean of the corresponding distribution. Taking the additional one-sided reach as \(C\) times this spread,  single mesh-scale layer to account for baseline damage, and identifying the resulting thickness with \(\ell_{fc}\) at the scaling level gives
\begin{equation}
\frac{\ell_{fc}}{\xi}\sim1+2C\frac{R_0}{\xi}\Delta_{\log}\sim1+C^\star\frac{\langle\ell\rangle}{\xi}\Delta_{\log}.
\label{eq:general_lfc_scaling}
\end{equation}
The factor two accounts for both sides of the crack plane. The reference coefficient is \(C^\star=2Cc_\ell\), where \(C\) is a distribution-dependent width factor. The derivation and width convention are detailed in the Supplemental Material.

For narrow reserve distributions it can be shown that \(\Delta_{\log}\simeq\sigma_\lambda/(\lambda_c-\langle\lambda\rangle)=\Delta^\star\), where \(\Delta^\star\) is the reserve coefficient of variation. Equation~\eqref{eq:general_lfc_scaling} then reduces to \(\ell_{fc}/\xi\sim1+C^\star(\langle\ell\rangle/\xi)\Delta^\star\), the approximation shown by the dashed curve in Fig.~\ref{fig:fractocohesive_length}(a). Broader reserve spread allows susceptible strands to rupture farther from the crack path, while the accompanying increase in the mean reference strand end-to-end distance enlarges the estimated deformation range, \(R_0\simeq c_\ell\langle\ell\rangle\), and amplifies this broadening. Through \(\Gamma=W_c\ell_{fc}\), the enlarged active region provides the spatial basis for the additional toughness contribution identified in Fig.~\ref{fig:introductory}(d).

    Finally, it is shown that fractocohesive length \(\ell_{fc}\) controls the crossover between flaw-insensitive and flaw-sensitive failure \cite{CHEN201750,Long2020TheFO}. Figure~\ref{fig:fractocohesive_length}(b) shows that work of fracture \(W_{\mathrm{frac}}\), defined as the area under the  stress--stretch curve of virgin sample up to the failure stretch of the corresponding notched sample, remains comparatively insensitive to notch length for \(c\lesssim\ell_{fc}\), but decreases markedly once \(c\gtrsim\ell_{fc}\). The larger \(\ell_{fc}\) of the polydisperse network ($\Delta^\star = 0.21$) therefore shifts the onset of flaw sensitivity to substantially larger defects. For the smallest notches, the monodisperse network can nevertheless exhibit a larger work of fracture because its more uniform load distribution delays localization in the near-defect-free limit \cite{Zhigang_perfect_vs_imperfect}. Thus, prestretch heterogeneity enhances flaw tolerance by broadening the crack-tip deformation field and extending the region over which strand scission occurs.

    Equation~\eqref{eq:general_lfc_scaling} suggests that increasing the logarithmic reserve width \(\Delta_{\log}\) can promote damage delocalization. This width can also increase through a contrast between the mean prestretches of two individually narrow populations. This motivates constructing bimodal networks in which highly prestretched strands can supply distributed rupture while a less prestretched matrix maintains load-bearing connectivity. We superimpose a dense, well-percolated matrix with mean reference end-to-end distance \(\langle\ell_1\rangle\) and a sparse sacrificial population with \(\langle\ell_2\rangle\) on the same crosslinking nodes. This construction directly controls \(\alpha=\langle\ell_2\rangle/\langle\ell_1\rangle\) and \(\langle\lambda_2\rangle/\langle\lambda_1\rangle\) which is analogous to double-network elastomer and hydrogel synthesis, in which a sacrificial network is swollen within the monomers of a secondary network \cite{Gong_2003_DN_Extreme_strength,Gong_DN_2010,slootman_pnas}.

\begin{figure}[t!]
    \centering
    \includegraphics[width=1.0\linewidth, trim={0.8cm 0.3cm 1.0cm 0cm}]{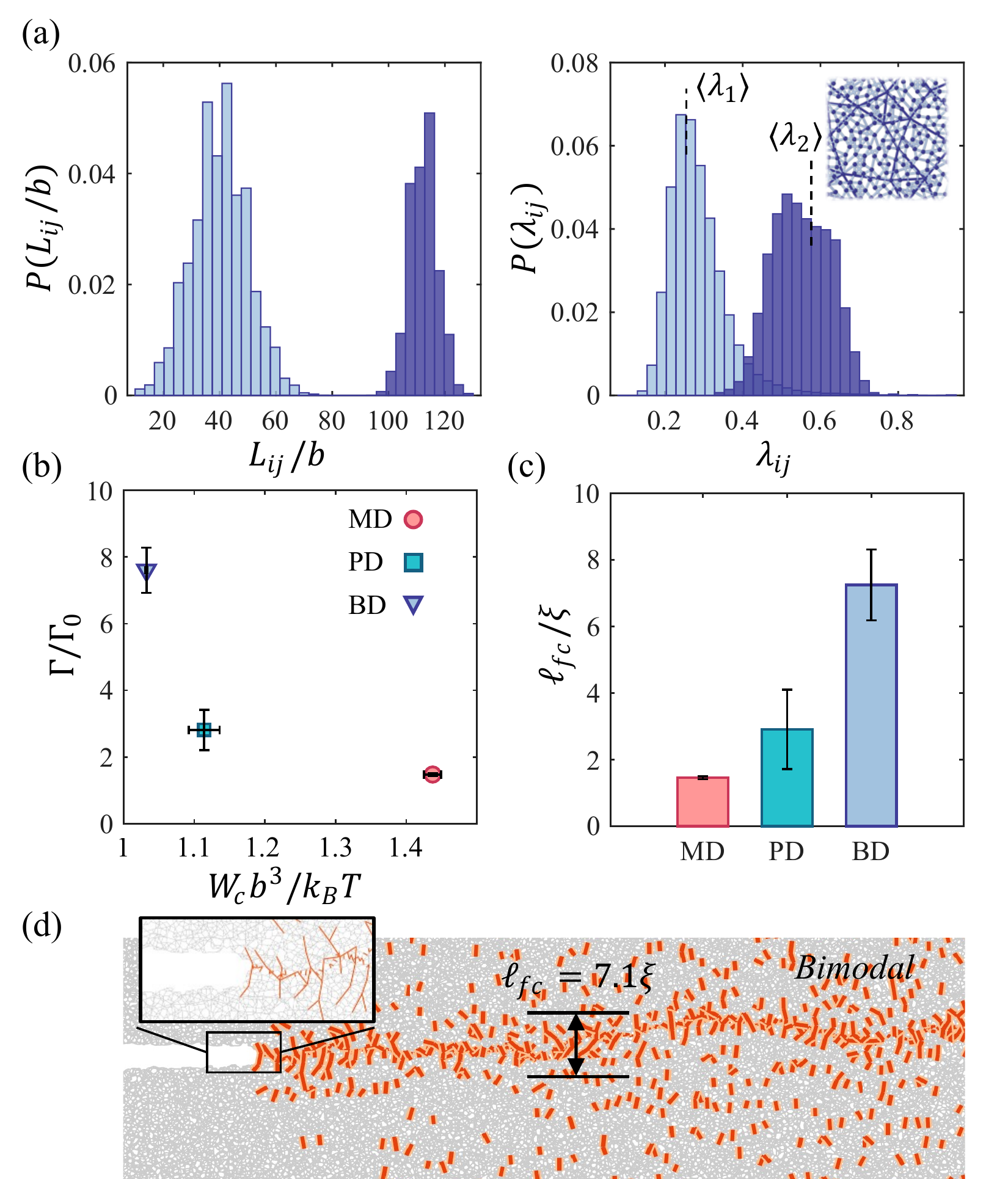}
    \caption{\footnotesize \justifying \textbf{Bimodal architecture.} (a) Normalized contour-length distributions \(P(L_{ij}/b)\) and equilibrated prestretch distributions \(P(\lambda_{ij})\) for a bimodal network, with inset showing the equilibrated snapshot. (b) Normalized fracture toughness \(\Gamma/\Gamma_0\) versus normalized critical work density \(W_c b^3/k_BT\) for monodisperse (MD), polydisperse (PD), and bimodal (BD) networks. (c) Fractocohesive length \(\ell_{fc}/\xi\) for three architectures. Polydisperse values in (b) and (c) are averaged over all \(\Delta^\star\) values. (d) Reference-configuration damage map for a bimodal network.}
    \label{fig:bimodal}
    \end{figure}

    Figure~\ref{fig:bimodal}(a) shows the resulting contour-length and prestretch distributions for a representative bimodal architecture with \(\alpha=5\) and \(\langle\lambda_2\rangle/\langle\lambda_1\rangle\approx2.5\). Two distinct contour-length populations are evident, corresponding to the short- and long-strand subnetworks. After equilibration, these map into a strongly bimodal prestretch distribution, in which the long strands are highly prestretched whereas the short strands remain comparatively weakly stretched. The inset in Fig.~\ref{fig:bimodal}(a) confirms the associated double-network-like morphology.

    The fracture consequences of this coupled heterogeneity are shown in Fig.~\ref{fig:bimodal}(b), using the same five independent realizations for each monodisperse or bimodal case and for each polydisperse disorder level. The reported polydisperse values are additionally averaged over all polydisperse \(\Delta^\star\) values studied. When the normalized toughness $\Gamma/\Gamma_0$ is plotted against the critical work density \(W_c\), the bimodal networks exhibit the largest toughness, reaching \(\Gamma/\Gamma_0\approx7.6\), despite not possessing the largest bulk energy-storage capacity. This suggests that enhanced fracture resistance is governed not simply by the defect-free critical work density, but by the spatial redistribution of bond rupture around the crack tip.

    This amplification is reflected directly in the fracture length scale. Figure~\ref{fig:bimodal}(c) shows that the fractocohesive length increases monotonically from monodisperse to polydisperse to bimodal architectures, reaching its largest value in the bimodal case. The corresponding damage map in Fig.~\ref{fig:bimodal}(d) reveals highly delocalized strand scission over a broad region, consistent with the energetic length \(\ell_{fc}\approx7.1\xi\). The inset further shows that this delocalization occurs predominantly in the highly prestretched long-strand (sacrificial) network, whereas damage in the short-strand network remains comparatively localized. This extensive delocalized damage is qualitatively consistent with the experimental observations \cite{slootman_pnas,Gong_2003_DN_Extreme_strength} on multiple network hydrogels and elastomers. We note, however, that the present model does not include entanglements in the short-strand network, which have been shown to further enlarge the fracture process zone, in some cases to several orders of magnitude beyond the mesh size \cite{Long2020TheFO,Junsoo_entangle_science}. Bimodal architectures therefore amplify toughness because strand length heterogeneity enhances the ability of prestretch heterogeneity to spread rupture away from the crack tip while preserving overall network connectivity.

    To examine the coupled effects of reserve spread and spatial range in Eq.~\eqref{eq:general_lfc_scaling}, we construct bimodal networks with prescribed mean prestretch ratios, \(\langle\lambda_2\rangle/\langle\lambda_1\rangle=1,\,1.5,\,2,\) and \(2.5\), and ratios of mean strand end-to-end distances, \(\alpha=\langle\ell_2\rangle/\langle\ell_1\rangle=1,\,2,\,3,\,4,\) and \(5\). Each parameter combination is averaged over three independent realizations. As defined above, the normalization length \(\xi\) is the mean nearest-neighbor crosslink distance of the complete equilibrated reference network.

\begin{figure}[t!]
    \centering
    \includegraphics[width=1.0\linewidth, trim={2cm 0.8cm 0.5cm 0cm}]{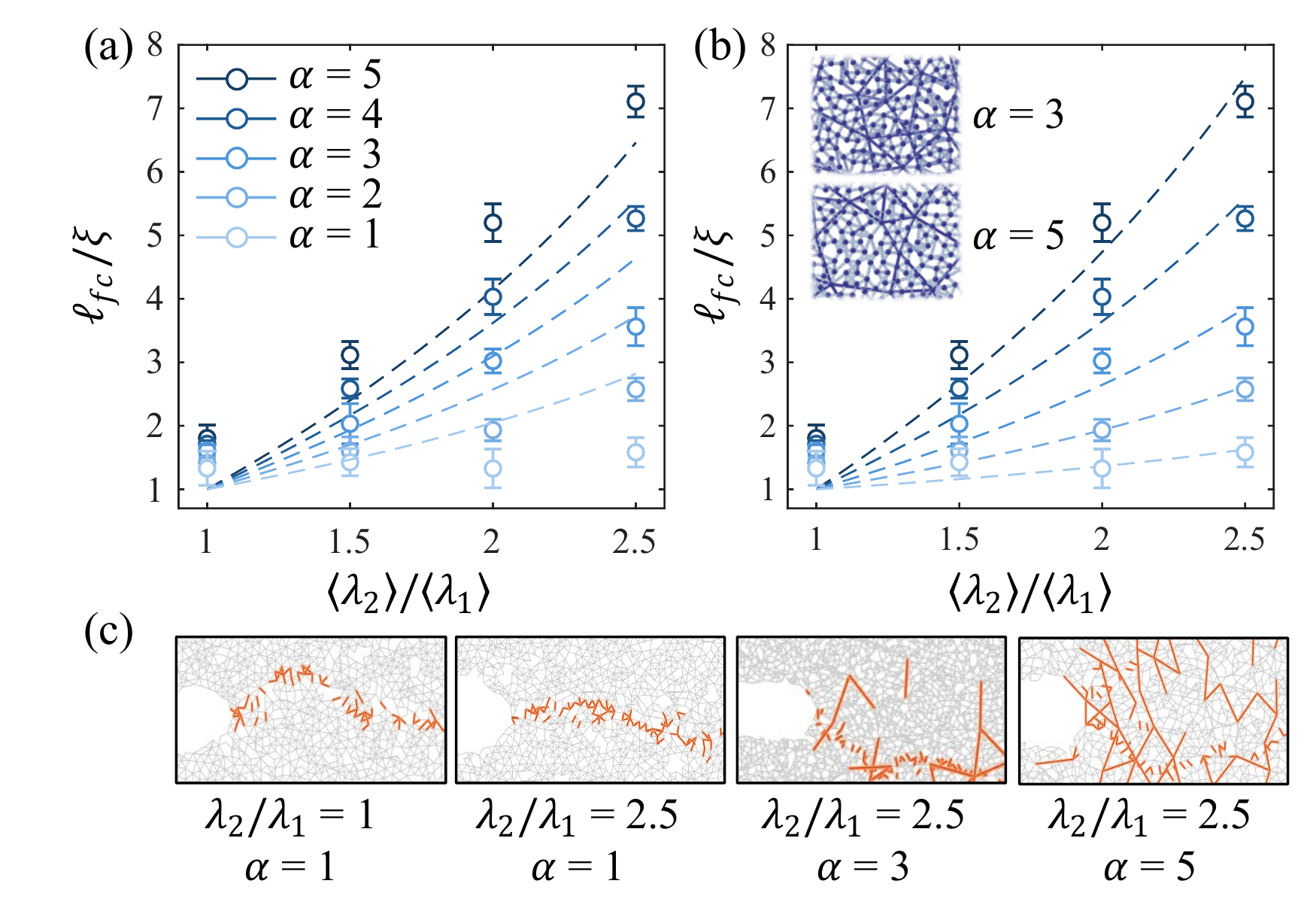}
    \caption{\footnotesize \justifying \textbf{Coupled heterogeneity in bimodal networks.} (a) Normalized fractocohesive length \(\ell_{fc}/\xi\) versus mean prestretch ratio \(\langle\lambda_2\rangle/\langle\lambda_1\rangle\) for different strand length ratios \(\alpha\). Dashed scaling curves are generated with fixed $c_\ell=1.54$ and $C=1.4007$. (b) The same simulation data compared with scaling curves using one effective coefficient $c_{\mathrm{eff}}$ fitted for each $\alpha$, while retaining the same $C$. (c) Reference-configuration damage maps.}
    \label{fig:phase_space}
    \end{figure}

Figure~\ref{fig:phase_space}(a) shows that increasing the prestretch ratio produces little delocalization when \(\alpha=1\), whereas the increase in fractocohesive length becomes progressively larger at greater geometric contrast. At \(\langle\lambda_2\rangle/\langle\lambda_1\rangle=2.5\), the normalized fractocohesive length increases from approximately \(1.6\xi\) to \(7.1\xi\). Additionally, as illustrated in figure~S2(a) the same trend appears in fracture toughness where at $\alpha=1$ the effect of prestretch is weak and increasing strand length ratio $\alpha$ amplifies the toughening. Thus, the largest fracture resistance occurs when prestretch contrast is combined with a contrast in the distances spanned by bonded strands.

The damage maps in Fig.~\ref{fig:phase_space}(c) show the accompanying redistribution of scission. At \(\alpha=1\), rupture remains concentrated near the crack path as the prestretch ratio increases from \(1\) to \(2.5\). However, at a prestretch ratio of \(2.5\), increasing \(\alpha\) to \(3\) and \(5\) produces progressively more rupture away from the crack plane. These maps connect the increase in the energetic fracture length to a broader region of strand scission. Prestretch-dependent damage delocalization has also been observed experimentally in multiple-network elastomers \cite{slootman_pnas}. Experiments on bimodal and trimodal PDMS networks prepared by end-linking mixtures of precursors with different molar masses, however, reported a more limited enhancement of fracture energy  \cite{Cohen_Genesky_bimodal}.

To interpret these trends, we approximate each prestretch population by its mean reserve, \(\bar\rho_s=\lambda_c-\langle\lambda_s\rangle\), with populations \(s=1,2\). The logarithmic reserve spread then becomes \(\Delta_{\log}\simeq\sqrt{w_1w_2}\left|\ln(\bar\rho_1/\bar\rho_2)\right|\), where \(w_s\) is the fraction of all intact reference strands belonging to \(s\), and \(w_1+w_2=1\). A larger separation between the mean reserves broadens the activation coordinates, while the factor \(\sqrt{w_1w_2}\) accounts for the relative abundance of the two populations.

The spatial range is estimated from the same relation \(R_0\simeq c_\ell\langle\ell\rangle\) used for mono- and polydisperse networks. For the bimodal architecture, the mean strand end-to-end distance is \(\langle\ell\rangle=w_1\langle\ell_1\rangle+w_2\langle\ell_2\rangle=\langle\ell_1\rangle[1+w_2(\alpha-1)]\). Consequently, \(R_0/\xi\simeq c_\ell(\langle\ell_1\rangle/\xi)[1+w_2(\alpha-1)]\), so the spatial amplification depends on both the strands span and their relative fraction. Equation~\eqref{eq:general_lfc_scaling} combines this architectural length with the logarithmic reserve spread to estimate the additional fracture-active thickness.

The dashed curves in Fig.~\ref{fig:phase_space}(a) evaluate Eq.~\eqref{eq:general_lfc_scaling} using the reference coefficients \(c_\ell=1.54\) and \(C=1.4007\), with the strand statistics specified in the Supplemental Material. While neglecting the internal reserve variations within each population the estimate captures the overall increase in fracture length but overpredicts the response at larger prestretch contrasts for the low-\(\alpha\) networks, particularly \(\alpha=1\) and \(2\). This discrepancy suggests that the mean strand span and reserve spread, combined through a fixed prefactor, do not fully capture the architecture dependence of damage broadening.

We account for this dependence through an effective coefficient \(c_{\mathrm{eff}}\), replacing \(c_\ell\) in the fracture-length estimate while retaining the same width convention and reserve statistics. A single \(c_{\mathrm{eff}}\) is fitted for each \(\alpha\) across the investigated prestretch ratios, corresponding to \(C^\star=2C c_{\mathrm{eff}}\). The resulting curves improve the agreement with the simulation data [Fig.~\ref{fig:phase_space}(b)]. Here, \(c_{\mathrm{eff}}\) describes the effective conversion of reserve heterogeneity and strand span into fracture-active thickness. Both estimates return to \(\ell_{fc}/\xi=1\) when the population mean prestretches coincide, so the observed offsets remain outside the narrow-population approximation.

Across the investigated bimodal family, the fitted ratio \(c_{\mathrm{eff}}/c_\ell\) increases with the strand-length heterogeneity measure \(\Delta_\ell\) [Fig.~S2(d)]. This trend is consistent with geometric heterogeneity helping sustain heterogeneous strand loading during deformation and rupture, allowing prestretch contrast to contribute more effectively to distributed damage. The proposed stabilization mechanism remains a physical interpretation of the observed architecture dependence.

    Together, these results show how equilibrium prestretch heterogeneity and network geometry jointly govern damage delocalization in polymer networks. The logarithmic reserve spread describes differences in susceptibility to rupture, while the mean strand end-to-end distance provides an architectural estimate of the range of crack-induced deformation. Their product provides the scaling variables for the additional fracture-active thickness, while an effective prefactor captures the architecture dependence of their conversion into distributed damage. The resulting length connects distributed scission to fracture resistance and flaw tolerance, with toughness also depending on the critical work density through \(\Gamma=W_c\ell_{fc}\). The reserve formulation also accommodates rupture-strength heterogeneity, as illustrated by the qualitative comparison in the Supplemental Material \cite{fulco2025fracturedisorderedstochasticlattice}. Three-dimensional simulations and explicit treatment of entanglements will be needed to assess how these trends extend quantitatively to experimental multiple network polymers\cite{Gong_2003_DN_Extreme_strength,slootman_pnas}.
    
    \textit{Acknowledgments---} We gratefully acknowledge the support of the 3M Company under sponsored research contract L-069903-L6-225 and National Science Foundation under award No. 2023179. Additionally, for this work, we utilized the Alpine high performance computing resource at the University of Colorado Boulder. Alpine is jointly funded by the University of Colorado Boulder, the University of Colorado Anschutz, Colorado State University, and the National Science Foundation (under award No. 2201538). Z.T.W. and F.J.V. acknowledge the support by the Department of Energy, National Nuclear Security Administration, Predictive Science Academic Alliance Program under Award Number DE-NA0003962. This report was prepared as an account of work sponsored by an agency of the United States Government. Neither the United States Government nor any agency thereof, nor any of their employees, makes any warranty, express or implied, or assumes any legal liability or responsibility for the accuracy, completeness, or usefulness of any information, apparatus, product, or process disclosed, or represents that its use would not infringe privately owned rights. Reference herein to any specific commercial product, process, or service by trade name, trademark, manufacturer, or otherwise does not necessarily constitute or imply its endorsement, recommendation, or favoring by the United States Government or any agency thereof. The views and opinions of authors expressed herein do not necessarily state or reflect those of the United States Government or any agency thereof.
    
    \bibliography{references.bib}
    
    \end{document}


\title{Supplemental Material for\\
\textit{Spreading the Damage: How Heterogeneity Delocalizes Fracture in Polymer Networks}}

\author{Saleh Assadi}
\author{Zachary White}
\author{Radhika Joshi}
\author{Franck J. Vernerey}
\email[correspondence to: ]{franck.vernerey@colorado.edu}

\affiliation{University of Colorado at Boulder, Boulder, Colorado, 80309, USA}

\maketitle
\renewcommand{\theequation}{S\arabic{equation}}
\renewcommand{\thefigure}{S\arabic{figure}}

\section*{Supplemental Material Overview}

This Supplemental Material provides the detailed formulation of the discrete mesoscale network model used in the main text.
We describe the nonlinear elasticity of polymer strands, nonbonded repulsive and cohesive interactions, and the numerical implementation employed in the simulations.
We also derive a scaling estimate that relates the logarithmic spread of rupture reserves across the complete strand distribution to the fracture-active length, and explain its evaluation for monodisperse, polydisperse, and bimodal networks and its extension to distributed rupture thresholds.


\subsection{Nonlinear elasticity of polymer chains}

In the discrete framework, each polymer strand connecting nodes $i$ and $j$ is modeled as a finitely extensible nonlinear spring.
With node positions $\bm{r}_i$ and $\bm{r}_j$, the end-to-end vector and its length are given by $\bm{r}_{ij} = \bm{r}_j - \bm{r}_i$ and $r_{ij} = \|\bm{r}_{ij}\|$, respectively.
A polymer strand comprising $n_{ij}$ Kuhn segments of length $b$ stores entropic free energy upon extension, which can be approximated by the Langevin–Padé model~\cite{Cohen1991}:
\begin{equation} \label{eq:langevin_energy}
    \mathcal{F}^{el}_{ij}(r_{ij})
    = k_B T \left[
        \frac{r_{ij}^2}{2 n_{ij} b^2}
        - n_{ij}\,\ln\!\left(\frac{n_{ij}^2 b^2 - r_{ij}^2}{n_{ij} b^2}\right)
    \right],
\end{equation}
where $k_B$ is Boltzmann’s constant and $T$ the absolute temperature.
The force on node $i$ due to strand $i\!-\!j$ follows from the derivative of the free energy with respect to $\bm{r}_i$,
\begin{equation}
    \bm{f}^{el}_{ij}
    \;=\; -\,\frac{\partial \mathcal{F}^{el}_{ij}}{\partial \bm{r}_i}
    \;=\; \frac{\partial \mathcal{F}^{el}_{ij}}{\partial \bm{r}_{ij}}
    \;=\; \frac{\mathrm{d}\mathcal{F}^{el}_{ij}}{\mathrm{d} r_{ij}}\,
          \frac{\bm{r}_{ij}}{r_{ij}}.
\end{equation}
Evaluating the radial derivative of Eq.~\ref{eq:langevin_energy} yields
\begin{equation}
\begin{aligned}    
    \bm{f}^{el}_{i j}
    = k_B T \left[
        \frac{1}{n_{ij} b^2}
        + \frac{2 n_{ij}}{\,n_{ij}^2 b^2 - r_{ij}^2\,}
    \right] \bm{r}_{ij}, \qquad
    \bm{f}^{el}_{j i} = -\,\bm{f}^{el}_{i j}.
\end{aligned}
\end{equation}
In the small-extension limit $r_{ij}\ll n_{ij} b$, this expression reduces to the Gaussian spring law
\[
\bm{f}^{el}_{i j}\approx \frac{3 k_B T}{n_{ij} b^2}\,\bm{r}_{ij}.
\]

\noindent To model irreversible strand scission, each bonded strand is assigned a rupture criterion based on its normalized extension
\begin{equation}
\lambda_{ij}=\frac{r_{ij}}{n_{ij}b}.
\end{equation}
A strand is irreversibly removed from the bonded network once
\begin{equation}
\lambda_{ij}\ge \lambda_c,
\label{eq:SI_rupture_rule}
\end{equation}
where \(\lambda_c=0.95\) is the critical rupture stretch used throughout this work. Upon rupture, the corresponding bonded interaction is deleted. Its rupture-energy contribution is \(E_{ij}^{\mathrm{rupture}}=n_{ij}U_b^f\), where \(U_b^f\) is the elastic energy per Kuhn segment accumulated between zero strand extension and the rupture threshold,
\begin{equation}
U_b^f=\frac{\mathcal{F}^{el}_{ij}(\lambda_c n_{ij}b)-\mathcal{F}^{el}_{ij}(0)}{n_{ij}}=k_BT\left[\frac{\lambda_c^2}{2}-\ln\!\left(1-\lambda_c^2\right)\right].
\label{eq:SI_rupture_energy}
\end{equation}
This energy reference gives the same \(U_b^f\) for every strand, while the rupture energy of each strand scales with its Kuhn-segment count.
\subsection{Non-bonded repulsive and cohesive interactions}

In addition to bonded polymer strands, non-bonded interactions between network nodes play a crucial role in maintaining network cohesion, enforcing near-incompressibility, and enabling realistic crack propagation.
To capture these effects, all nodes, whether bonded or not, and their local environments interact through two complementary energetic contributions: a \textit{soft repulsive pair potential} that prevents node overlap and a \textit{cohesive multibody potential} based on local density that captures the attractive forces sustaining the network.\\

\noindent To prevent unphysical overlap of nodes, a short-range repulsive potential is introduced between all  pairs within a cutoff distance $r_c$.
This soft potential combines harmonic and anharmonic terms to produce a smooth yet strongly repulsive response:
\begin{equation}
\begin{aligned}
\mathcal{F}^{\mathrm{rep}}_{ij}(r_{ij})
= \frac{\epsilon_{ij}}{2}(r_{ij}-\sigma_{ij})^2
 + \frac{\epsilon^a_{ij}}{4}(r_{ij}-\sigma_{ij})^4, 
\qquad r_{ij} \le r_c ,
\end{aligned}
\end{equation}
where $\epsilon_{ij}$ and $\epsilon^a_{ij}$ denote the harmonic and anharmonic stiffnesses, respectively, and $\sigma_{ij}$ defines the preferred equilibrium spacing between nodes. In order to only induce repulsion, we set the cutoff radius $r_c=\sigma_{ij}$ equal to equilibrium spacing.
The corresponding repulsive force is
\begin{equation}
\begin{aligned}
\bm{f}^{\mathrm{rep}}_{ij}
= -\,\frac{\partial \mathcal{F}^{\mathrm{rep}}_{ij}}{\partial \bm{r}_i} 
= \Bigl[\epsilon_{ij}(r_{ij}-\sigma_{ij})
        + \epsilon^a_{ij}(r_{ij}-\sigma_{ij})^3\Bigr]
        \frac{\bm{r}_{ij}}{r_{ij}}, \qquad
\bm{f}^{\mathrm{rep}}_{ji}
= -\bm{f}^{\mathrm{rep}}_{ij}.
\end{aligned}
\end{equation}                                                                                  
The anharmonic contribution amplifies the stiffness at small separations, effectively regularizing the excluded-volume interaction while maintaining smooth force continuity across the cutoff distance.\\

\noindent The effective length scale $\sigma_{ij}$ governs the equilibrium separation between two non-bonded nodes and is linked to the average polymer density of the network.
Specifically, $\sigma_{ij}$ is set proportional to the mean volume per node associated with the total Kuhn segments in the system:
\[
\sigma_{ij} \;\propto\;
\left(
\frac{N^{\text{Kuhn}}_{\text{tot}}}{N_{\text{nodes}}}
\right)^{1/2} b,
\]
where $N^{\text{Kuhn}}_{\text{tot}} = \sum_{\langle i,j\rangle} n_{ij}$ is the total number of Kuhn segments in all bonded strands, $N_{\text{nodes}}$ is the total number of nodes in the network, and $b$ is the Kuhn length.
This coarse-grained scaling connects the non-bonded interaction range to the global polymer density, effectively defining the equilibrium spacing between network junctions.
By incorporating this density-dependent length scale, the repulsive potential introduces a finite resistance to volumetric collapse and ensures consistent scaling across different network densities.\\

\noindent Complementing this excluded-volume interaction, a cohesive multibody potential captures the long-range attraction between nodes and governs the macroscopic cohesion of the network.
Following the framework of the pair style local/density potential in LAMMPS \cite{Sanyal16,Sanyal18}, each node $i$ is assigned a local density $\rho_i$, computed as a weighted sum of the contributions from neighboring nodes through a smooth kernel function $\varphi(r_{ij})$,
\begin{equation}
    \rho_i = \sum_{j\neq i} \varphi(r_{ij}).
\end{equation}
The kernel $\varphi(r_{ij})$ acts within a finite interaction range defined by two cutoffs: a lower cutoff $r_1$ and an upper cutoff $r_2$.
It is constant ($\varphi=1$) for $r_{ij} < r_1$, decays smoothly to zero for $r_{ij} > r_2$, and is twice continuously differentiable within the transition region $r_1 < r_{ij} < r_2$.
The cutoffs are scaled relative to the equilibrium spacing $\sigma_{ij}$ as
\[
r_1 = 0.8\,\sigma_{ij}, 
\qquad
r_2 = 4\,\sigma_{ij}.
\]
Under this construction, the equilibrium local density scales with the cutoff ratio as
\begin{equation}
    \rho_0 \propto \left(\frac{r_2}{\sigma_{ij}}\right)^{2},
\end{equation}
which in the present system yields $\rho_0 \approx 16$.\\

\noindent
The cohesive energy associated with each node is then defined as
\begin{equation}
    \mathcal{F}^{\mathrm{coh}}_i
    = \epsilon_{LD}\,(\rho_i - \rho_0)^2,
\end{equation}
where $\epsilon_{LD}$ denotes the cohesive strength and $\rho_0$ is the equilibrium local density.
This harmonic form penalizes deviations from equilibrium packing, thereby introducing an effective volumetric stiffness that stabilizes the network under deformation and provides cohesive tractions near crack surfaces.
The corresponding pairwise cohesive force is obtained from the gradient of the cohesive potential as
\begin{equation}
\bm{f}^{\mathrm{coh}}_{ij}
= {} - \frac{\partial \mathcal{F}^{\mathrm{coh}}_{i}}{ \partial \bm{r}_i}  
= 2\epsilon_{LD}
      \bigl[\rho_i + \rho_j - 2\rho_0 \bigr]\,
      \frac{1}{\langle\varphi\rangle}
      \frac{d\varphi(r_{ij})}{d r_{ij}}
      \frac{\bm{r}_{ij}}{r_{ij}},
\end{equation}
This expression naturally introduces cohesive tractions that depend on the local packing state and smoothly vanish when the local density approaches its reference value, ensuring energy consistency and momentum conservation. In the present simulations, the local-density kernel is taken in the piecewise form
\begin{equation}
\varphi(r)=
\begin{cases}
1, & r\le r_1, \\[4pt]
c_0+c_2r^2+c_4r^4+c_6r^6, & r_1<r<r_2, \\[4pt]
0, & r\ge r_2,
\end{cases}
\label{eq:SI_kernel}
\end{equation}
where the coefficients \(c_0,c_2,c_4,c_6\) are chosen so that \(\varphi(r)\) and its derivative remain continuous at the matching points \(r_0\) and \(r_c\). The lower and upper cutoffs correspond to the inner flat region and the outer interaction range of the local-density potential, respectively.\\

\noindent To avoid artificial overcounting of local volume, the kernel is normalized by its spatial average over the interaction domain. Accordingly, the density entering the cohesive potential is computed as
\begin{equation}
\rho_i
=
\sum_{j\neq i}
\frac{\varphi(r_{ij})}{\langle \varphi \rangle},
\label{eq:SI_rho_normalized}
\end{equation}
where \(\langle \varphi \rangle\) denotes the average kernel weight over the cutoff region. This normalization ensures that the computed local density reflects the effective packing state rather than the absolute integrated weight of the polynomial kernel.

\section{Numerical implementation}

All simulations are performed in LAMMPS \cite{THOMPSON2022108171} under quasistatic loading conditions. Following network generation, node positions are relaxed using overdamped dynamics until force balance is achieved. The nodal equations of motion are integrated using the standard velocity-Verlet scheme (`fix nve`) together with a viscous drag term (`fix viscous`), so that the motion of node \(i\) obeys
\begin{equation}
m_i \ddot{\bm r}_i
=
\bm f_i^{\mathrm{tot}}
-
\eta \dot{\bm r}_i,
\label{eq:SI_damped_eom}
\end{equation}
where \(m_i\) is the nodal mass, \(\eta\) is the viscous damping coefficient, and \(\bm f_i^{\mathrm{tot}}\) is the total force acting on the node. This total force includes bonded elastic forces and non-bonded repulsive and cohesive interactions,
\begin{equation}
\bm f_i^{\mathrm{tot}}
=
\sum_{j\in\mathcal{B}_i} \bm f_{ij}^{\mathrm{el}}
+
\sum_{j\neq i}\left(\bm f_{ij}^{\mathrm{rep}}+\bm f_{ij}^{\mathrm{coh}}\right).
\label{eq:SI_total_force}
\end{equation}
The damping is chosen sufficiently large that inertial effects remain negligible and the dynamics operate close to the overdamped limit,
\begin{equation}
\eta \dot{\bm r}_i \approx \bm f_i^{\mathrm{tot}}.
\label{eq:SI_overdamped}
\end{equation}

\noindent After equilibration, uniaxial deformation is applied at a constant engineering strain rate,
\begin{equation}
\dot{\epsilon}=\frac{\dot L_y}{L_y^0},
\end{equation}
where \(L_y^0\) is the initial box length in the loading direction. The strain rate is chosen sufficiently small to ensure near-equilibrium response throughout the loading process. Quasistatic conditions are verified by monitoring the ratio of total kinetic to total potential energy, which remains negligible throughout deformation, ensuring that inertial effects do not influence stress transmission, bond rupture, or crack propagation. Specifically,
\begin{equation}
\frac{K}{U} < 10^{-5}
\end{equation}
at all times, and further reductions in strain rate do not alter the stress--stretch response or the extracted fracture metrics.

\begin{table}[h!]
\caption{\label{tab:metrics}
Definitions of the principal stress and fracture metrics used in the main text.}
\begin{ruledtabular}
\begin{tabular}{lll}
Symbol & Expression & Physical meaning \\
\hline
\(\sigma_{\alpha\beta}\) &
\(\displaystyle \frac{1}{V}\sum_i\sum_{j>i} r_{ij,\alpha}f_{ij,\beta}\) &
Configurational virial stress \\[6pt]

\(W_{\mathrm{frac}}\) &
\(\displaystyle \int_{1}^{\lambda_f^{\mathrm{notch}}}\sigma_{\mathrm{virgin}}(\lambda)\,d\lambda\) &
Work of fracture of notched sample \\[8pt]

\(W_c\) &
\(\displaystyle \int_{1}^{\lambda_f^{\mathrm{virgin}}}\sigma_{\mathrm{virgin}}(\lambda)\,d\lambda\) &
Critical work density of virgin sample \\[8pt]

\(\Gamma\) &
\(\displaystyle \frac{1}{A_{\mathrm{crack}}}\sum_{k\in\mathcal{R}} E_k^{\mathrm{rupture}}\) &
Fracture toughness from strand rupture \\[8pt]

\(\Gamma_0\) &
\(\displaystyle \nu \langle n\rangle U_b^f \xi\) &
Lake--Thomas reference fracture scale \\[6pt]

\(\ell_{fc}\) &
\(\displaystyle \frac{\Gamma}{W_c}\) &
Fractocohesive length \\[8pt]

\(\Delta^\star\) & \(\displaystyle \frac{\sigma_\lambda}{\lambda_c-\langle\lambda\rangle}\) & Reserve coefficient of variation \\[8pt]
\(\Delta_{\log}\) & \(\displaystyle \operatorname{std}_{\mathrm{all}}(\ln\rho)\) & Logarithmic reserve spread \\[8pt]

\(\alpha\) &
\(\displaystyle \frac{\langle\ell_2\rangle}{\langle\ell_1\rangle}\) &
equilibrium strand length ratio in bimodal networks \\[8pt]

\end{tabular}
\end{ruledtabular}
\end{table}

\section{Stress and fracture metrics}

To quantify the mechanical and fracture response of the network, we compute stress, work densities, and fracture metrics directly from the simulated force--displacement response of virgin and notched samples.\\

\noindent The stress reported in this work is obtained from the configurational virial in LAMMPS using `compute pressure`, with the kinetic (thermal) contribution omitted. Since the simulations are performed under overdamped, quasistatic conditions with fixed boundaries, the stress reduces to the potential part of the virial,
\begin{equation}
\sigma_{\alpha\beta}
=
\frac{1}{V}
\sum_{i}
\sum_{j>i}
r_{ij,\alpha}\,f_{ij,\beta},
\label{eq:SI_virial}
\end{equation}
where \(V\) is the system volume (or area times unit thickness in the present 2D setting), \(r_{ij,\alpha}\) is the \(\alpha\)-component of the inter-nodal separation vector, and \(f_{ij,\beta}\) is the \(\beta\)-component of the interaction force between nodes \(i\) and \(j\). The tensile stress \(\sigma\) shown in the main text and Supplemental Material corresponds to the loading-direction component of this virial stress.\\

\noindent The work of fracture \(W_{\mathrm{frac}}\) is defined from the stress--stretch response of a notched sample as the area under the virgin stress--stretch curve up to the failure stretch of the corresponding notched sample,
\begin{equation}
W_{\mathrm{frac}}
=
\int_{1}^{\lambda_f^{\mathrm{notch}}}
\sigma_{\mathrm{virgin}}(\lambda)\,d\lambda,
\label{eq:SI_Wfrac}
\end{equation}
where \(\lambda_f^{\mathrm{notch}}\) denotes the stretch at which the notched sample loses load-bearing capacity.\\

\noindent The critical work density \(W_c\) is obtained from the unnotched stress--stretch response up to the failure or localization point of the virgin sample,
\begin{equation}
W_c
=
\int_{1}^{\lambda_f^{\mathrm{virgin}}}
\sigma_{\mathrm{virgin}}(\lambda)\,d\lambda,
\label{eq:SI_Wc}
\end{equation}
where \(\lambda_f^{\mathrm{virgin}}\) denotes the critical stretch of the virgin sample used in the main text.\\

\noindent The fracture toughness \(\Gamma\) is computed directly from the total bonded elastic energy released by irreversible strand rupture during crack advance, divided by the corresponding crack extension area,
\begin{equation}
\Gamma
=
\frac{1}{A_{\mathrm{crack}}}
\sum_{k\in \mathcal{R}} E_{k}^{\mathrm{rupture}},
\label{eq:SI_Gamma}
\end{equation}
where \(\mathcal{R}\) denotes the set of ruptured strands and \(E_k^{\mathrm{rupture}}\) is the bonded elastic energy stored in strand \(k\) at the instant of rupture. In the present 2D setting, \(A_{\mathrm{crack}}\) is the crack extension length multiplied by the unit out-of-plane thickness equal to Kuhn length $b$.\\

\noindent The Lake--Thomas reference fracture scale is written as
\begin{equation}
\Gamma_0 \sim \nu \langle n\rangle U_b^f \xi,
\label{eq:SI_Gamma0}
\end{equation}
where \(\nu\) is the crosslink density, \(\langle n\rangle\) is the average Kuhn-segment count per strand, \(U_b^f\) is the rupture energy per Kuhn segment, and \(\xi\) is the mesh size measured as the average intercrosslinker distance within the first coordination shell of the equilibrated reference network.\\

\noindent The fractocohesive length is then defined as
\begin{equation}
\ell_{fc}=\frac{\Gamma}{W_c},
\label{eq:SI_lfc}
\end{equation}
which provides the intrinsic fracture length scale used throughout the main text to characterize the spatial extent of the damage process zone and the crossover between flaw-insensitive and flaw-sensitive response.

\section{Scaling estimate for the damage zone lengthscale}

\begin{figure}[ht!]
\centering
\includegraphics[width=0.98\linewidth]{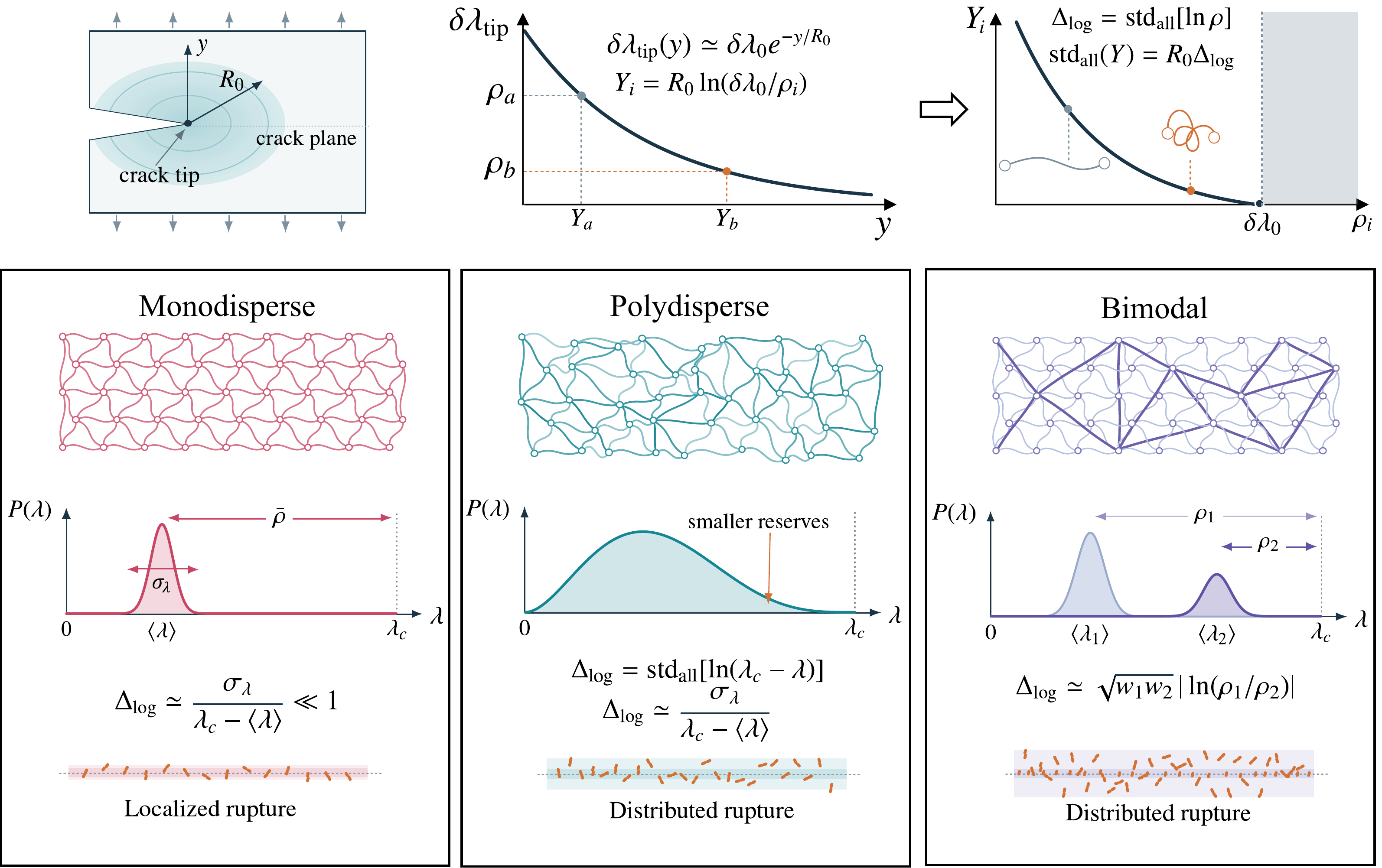}
\caption{\textbf{Rupture reserves and damage broadening.} The upper row illustrates the crack-induced stretch field, its intersections with different reserves, and the logarithmic reserve-to-distance mapping. The shaded range $\rho>\delta\lambda_0$ is inaccessible at the illustrated amplitude. The lower row shows how narrow, broad, and bimodal prestretch distributions produce different reserve spreads and corresponding rupture patterns. The approximation $\Delta_{\log}\simeq\sigma_\lambda/(\lambda_c-\langle\lambda\rangle)$ applies to narrow distributions, while the bimodal expression applies to two narrow populations with strand fractions $w_1$ and $w_2$. The field range is estimated from the mean bond end-to-end distance through $R_0\simeq c_\ell\langle\ell\rangle$, with $c_\ell=1.54$.}
\label{fig:scaling_law_schematic}
\end{figure}

Prestretch heterogeneity gives strands different remaining stretches to rupture, allowing a decaying crack-induced stretch field to activate rupture over a range of distances from the crack plane. We relate this spatial broadening to the spread of rupture reserves and the mean bond end-to-end distance, and estimate the resulting fracture-active thickness, as illustrated in Fig.~\ref{fig:scaling_law_schematic}.

\subsection{Rupture reserves and the crack-induced stretch field}

Consider a crack propagating along the $x$-direction, with transverse distance $y\geq0$ measured from the crack plane in the reference configuration. An intact strand $i-j$ with equilibrium prestretch $\lambda_{ij}$ has rupture reserve
\begin{equation}
\rho_{ij}=\lambda_c-\lambda_{ij}>0, \qquad \bar\rho=\langle\rho\rangle=\lambda_c-\langle\lambda\rangle, \qquad \sigma_\rho=\sigma_\lambda,
\label{eq:SI_population_reserve}
\end{equation}
where $\lambda_c=0.95$ is the common rupture threshold and the brackets denote an equally weighted average over all intact strands in the reference network. Because the threshold is shared, reserve and prestretch fluctuations have the same magnitude, so
\begin{equation}
\Delta^\star=\frac{\sigma_\rho}{\bar\rho}=\frac{\sigma_\lambda}{\lambda_c-\langle\lambda\rangle}
\label{eq:SI_population_width}
\end{equation}
is the coefficient of variation of the rupture-reserve distribution, with $\rho$ distinct from the packing density used in the cohesive model. This spread describes differences in remaining stretch among strands, each of which still ruptures deterministically when its actual stretch reaches $\lambda_c$.\\

\noindent To describe how the crack activates rupture of strands away from its plane, we approximate the effective additional stretch by a common exponential envelope,
\begin{equation}
\delta\lambda_{\mathrm{tip}}(y)\simeq\delta\lambda_0\exp(-y/R_0), \qquad \delta\lambda_0>0,\quad R_0>0,
\label{eq:SI_tip_envelope}
\end{equation}
where $\delta\lambda_0$ is the stretch amplitude at the crack tip and $R_0$ is the decay length. A strand's reserve can be reached wherever this increment is sufficiently large, and equating the two gives its limiting activation coordinate,
\begin{equation}
\delta\lambda_0e^{-Y_{ij}/R_0}=\rho_{ij} \quad\Longrightarrow\quad Y_{ij}=R_0\ln\!\left(\frac{\delta\lambda_0}{\rho_{ij}}\right).
\label{eq:SI_general_boundary}
\end{equation}
For $0<\rho_{ij}\leq\delta\lambda_0$, $Y_{ij}$ is the largest transverse distance at which the field can reach the reserve, so a highly prestretched strand can be activated farther from the crack plane than a less stretched strand. Strands with $\rho_{ij}>\delta\lambda_0$ are inaccessible even at the crack plane and have negative formal coordinates. The equilibrium prestretches provide reference-state estimates of the rupture reserves, while the decay length measured ahead of the crack provides an estimate of the transverse range in Eq.~\eqref{eq:SI_tip_envelope}. A description relative to the loaded background would instead pair the reserves $\lambda_c-\lambda_{\mathrm{bg},ij}$ with the crack-induced stretch in excess of that background.

\subsection{Mean bond length and the crack-induced deformation range}

Let $\mathcal{B}$ denote the set of all intact bonded strands in the equilibrated reference network, with each strand counted once, and let $N_b$ be their total number. The mean bond end-to-end distance is
\begin{equation}
\langle\ell\rangle=\frac{1}{N_b}\sum_{(i,j)\in\mathcal{B}}r_{ij}^{\mathrm{ref}},
\label{eq:SI_mean_bond_length}
\end{equation}
where $r_{ij}^{\mathrm{ref}}$ is the reference end-to-end distance of strand $i-j$. This average includes all bonded connections. The mesh size $\xi$ is the average intercrosslinker distance within the first coordination shell. Thus, $\xi$ measures local crosslink spacing, while $\langle\ell\rangle$ measures the mean distance spanned by the network strands.\\

\noindent The approximately linear relation between $R_0/\xi$ and $\langle\ell\rangle/\xi$ in the inset of Fig.~2(d) of the main text motivates the architectural estimate
\begin{equation}
R_0\simeq c_\ell\langle\ell\rangle, \qquad c_\ell=1.54.
\label{eq:SI_bond_range}
\end{equation}
We use the same empirical coefficient $c_\ell$ for monodisperse, polydisperse, and bimodal networks. The mean bond length therefore sets one common field range for the complete network, with architectural differences entering through the measured $\langle\ell\rangle$.

\subsection{Logarithmic reserve spread and fracture-active length}

The logarithmic mapping in Eq.~\eqref{eq:SI_general_boundary} converts reserve distribution into spatial separations, with the common amplitude shifting all activation coordinates by the same amount. Averaging over the complete strand list and subtracting the mean therefore isolates the differences caused by reserve heterogeneity,
\begin{equation}
\langle Y\rangle=R_0\ln\delta\lambda_0-R_0\langle\ln\rho\rangle, \qquad Y_{ij}-\langle Y\rangle=-R_0\left(\ln\rho_{ij}-\langle\ln\rho\rangle\right),
\label{eq:SI_centered_boundary}
\end{equation}
so the variance of the activation coordinates is
\begin{equation}
\left\langle\left(Y-\langle Y\rangle\right)^2\right\rangle=R_0^2\left\langle\left(\ln\rho-\langle\ln\rho\rangle\right)^2\right\rangle.
\label{eq:SI_boundary_variance}
\end{equation}
This relation motivates the dimensionless logarithmic reserve width,
\begin{equation}
\Delta_{\log}=\left\langle\left(\ln\rho-\langle\ln\rho\rangle\right)^2\right\rangle^{1/2}=\operatorname{std}_{\mathrm{all}}(\ln\rho), \qquad \operatorname{std}_{\mathrm{all}}(Y)=R_0\Delta_{\log}.
\label{eq:SI_log_reserve_width}
\end{equation}
Thus, $\Delta_{\log}$ measures the spread of rupture reserves on a logarithmic scale, while $R_0$ converts that spread into a length. The relation is exact for the untruncated coordinates within the assumed common envelope, because changing the amplitude shifts their mean without changing their spread. Accessibility still depends on the amplitude, so restricting the average to strands with $Y\geq0$ would generally give a different, amplitude-dependent width.\\

\noindent To estimate a representative active thickness, we take the additional one-sided reach to be $C\operatorname{std}_{\mathrm{all}}(Y)$ and add a localized mesh-scale layer of thickness $\xi$. Including both sides of the crack plane and identifying the resulting thickness with the energetic length $\ell_{fc}=\Gamma/W_c$ at the scaling level gives
\begin{equation}
\boxed{\frac{\ell_{fc}}{\xi}\sim1+2C\frac{R_0}{\xi}\Delta_{\log}.}
\label{eq:SI_unified_scaling}
\end{equation}
We use $C=\sqrt{2/\pi}+\sqrt{1-2/\pi}\simeq1.4007$, the mean plus one standard deviation of the positive half of a centered Gaussian of unit variance, as a fixed width convention across all architectures. This choice sets the conversion from spread to representative reach without requiring the actual reserve distribution to be Gaussian, while the identification with $\ell_{fc}$ remains a physical scaling assumption for comparable states of active fracture.\\

\noindent Combining Eqs.~\eqref{eq:SI_bond_range} and \eqref{eq:SI_unified_scaling} gives the bond-length form of the scaling,
\begin{equation}
\boxed{\frac{\ell_{fc}}{\xi}\sim1+2Cc_\ell\frac{\langle\ell\rangle}{\xi}\Delta_{\log}=1+C^\star\frac{\langle\ell\rangle}{\xi}\Delta_{\log}.}
\label{eq:SI_bond_length_scaling}
\end{equation}
The constant $C^\star = 2 C c_{\ell}$ is introduced for cleaner scaling notation. Here $c_\ell=1.54$ sets the relation between mean bond length and field range, while $C=1.4007$ converts the activation-coordinate spread into a representative one-sided reach. Both coefficients are held fixed across architectures, giving $2Cc_\ell\simeq4.31$. The ratio $\langle\ell\rangle/\xi$ accounts for the spatial range of the bonded connections, while $\Delta_{\log}$ accounts for their spread in rupture reserves.\\

\noindent The enlarged active region contributes to fracture resistance through the energetic definition of $\ell_{fc}$, which retains the critical work density when converting the length estimate into toughness,
\begin{equation}
\frac{\Gamma}{\Gamma_0}=\frac{W_c\xi}{\Gamma_0}\frac{\ell_{fc}}{\xi}\sim\frac{W_c\xi}{\Gamma_0}\left[1+C^\star\frac{\langle\ell\rangle}{\xi}\Delta_{\log}\right].
\label{eq:SI_energy_length_connection}
\end{equation}
The prefactor $W_c\xi/\Gamma_0$ therefore carries the accompanying changes in bulk work density and the reference fracture scale, which also depends on strand density and rupture energy.

\subsection{Mono- and polydisperse networks}

For a single population with a narrow prestretch distribution, the fluctuations in reserve are small compared with its mean. Writing $\lambda=\langle\lambda\rangle+\delta\lambda$ and $\bar\rho=\lambda_c-\langle\lambda\rangle$, we expand for $|\delta\lambda|/\bar\rho\ll1$ to obtain
\begin{equation}
\ln\rho=\ln\bar\rho+\ln\!\left(1-\frac{\delta\lambda}{\bar\rho}\right)\simeq\ln\bar\rho-\frac{\delta\lambda}{\bar\rho},
\label{eq:SI_log_expansion}
\end{equation}
so the logarithmic width reduces to the coefficient of variation and gives the narrow-distribution form of Eq.~\eqref{eq:SI_bond_length_scaling},
\begin{equation}
\Delta_{\log}\simeq\frac{\sigma_\lambda}{\bar\rho}=\Delta^\star, \qquad \frac{\ell_{fc}}{\xi}\sim1+C^\star\frac{\langle\ell\rangle}{\xi}\Delta^\star.
\label{eq:SI_single_population_scaling}
\end{equation}
For broader distributions, evaluating $\Delta_{\log}$ directly retains the nonlinear sensitivity to strands close to rupture, whereas the approximation by $\Delta^\star$ neglects this effect. Even a monodisperse contour-length network can have a small nonzero reserve width because its equilibrated strand geometry is heterogeneous.\\

\noindent Table~\ref{tab:lfc_scaling_comparison} gives a numerical comparison of Eq.~\eqref{eq:SI_bond_length_scaling} for the smallest- and largest-disorder cases using the reported prestretch statistics and measured average strand lengths. The log widths in this example are reconstructed from the reference strand reserve distribution, which requires no assumed distribution shape. This comparison uses the measured $\langle \ell \rangle$ directly. 

\begin{table}[h!]
\centering
\begin{tabular}{lccccccc}
\hline
Network & $\langle\lambda\rangle$ & $\sigma_\lambda$ & $\Delta^\star$ & $\Delta_{\log}$ & $\langle\ell\rangle/\xi$ & $\ell_{fc}/\xi$ & $\ell_{fc}^{\mathrm{pred}}/\xi$\\
\hline
Mono & 0.23 & 0.029 & 0.0403 & 0.0404 & 1.1267 & 1.3 & 1.196\\
Poly & 0.40 & 0.120 & 0.2182 & 0.2346 & 2.5233 & 4.1 & 3.554\\
\hline
\end{tabular}
\caption{Measured fractocohesive lengths and bond-length scaling estimates for the monodisperse and most heterogeneous polydisperse networks. Predictions use Eq.~\eqref{eq:SI_bond_length_scaling} with fixed $c_\ell=1.54$ and $C=1.4007$, the unrounded mean values of $\langle\ell\rangle/\xi$ over three realizations, and the listed logarithmic reserve widths.}
\label{tab:lfc_scaling_comparison}
\end{table}

\subsection{Bimodal and multiple-network architectures}

Consider first two narrow prestretch populations, so that strands within each population have approximately the same reserve $\bar\rho_s=\lambda_c-\langle\lambda_s\rangle$, with $s=1,2$. In the present bimodal networks, the less prestretched matrix forms population 1, while the more highly prestretched sacrificial strands form population 2 and have the smaller reserve. These two reserves therefore correspond to different activation distances under the same crack-induced stretch field.\\
\noindent If population $s$ contains $N_{b,s}$ intact reference strands, its fraction is $w_s=N_{b,s}/N_b$, where $N_b=N_{b,1}+N_{b,2}$ and $w_1+w_2=1$. Their mean log reserve is $\langle\ln\rho\rangle\simeq w_1\ln\bar\rho_1+w_2\ln\bar\rho_2$. Each population contributes its fraction of the squared deviations from this mean, giving
\begin{equation}
\begin{aligned}
\Delta_{\log}^2
\simeq
w_1\left(\ln\bar\rho_1-\langle\ln\rho\rangle\right)^2
+w_2\left(\ln\bar\rho_2-\langle\ln\rho\rangle\right)^2 
=
w_1w_2\left[\ln\!\left(\frac{\bar\rho_1}{\bar\rho_2}\right)\right]^2.
\end{aligned}
\label{eq:SI_bimodal_log_variance}
\end{equation}
Taking the square root makes the contribution of the reserve contrast explicit,
\begin{equation}
\Delta_{\log}\simeq
\sqrt{w_1w_2}
\left|\ln\!\left(\frac{\bar\rho_1}{\bar\rho_2}\right)\right|.
\label{eq:SI_bimodal_narrow}
\end{equation}
Mean prestretch contrast can therefore broaden the activation coordinates even when each population has negligible internal width. The factor $\sqrt{w_1w_2}$ accounts for strand abundance, so the same contrast contributes less to the whole-network spread when one population is sparse, with both weights determined by the actual strand counts.\\
\noindent When the populations have finite widths, $\Delta_{\log}$ is evaluated over the full strand list and includes the variations within each subnetwork as well as the contrast between them. More generally, for any number of populations with strand fractions $w_s$, mean log reserves $\mu_s=\langle\ln\rho\rangle_s$, and log-reserve variances $v_s=\operatorname{Var}_s(\ln\rho)$,
\begin{equation}
\Delta_{\log}^2
=
\sum_s w_s v_s
+\sum_s w_s\left(\mu_s-\langle\ln\rho\rangle\right)^2,
\qquad
\langle\ln\rho\rangle=\sum_s w_s\mu_s.
\label{eq:SI_total_log_variance}
\end{equation}
The first term describes internal reserve heterogeneity and the second describes separation between populations, while their sum remains the same whole-network width regardless of how strands are grouped.\\
\noindent The reserve spread determines the differences in susceptibility to rupture, while the field range sets how far these differences can extend damage from the crack plane. Let $\langle\ell_s\rangle$ denote the mean strand end-to-end distance within population $s$ in the equilibrated reference network, with each intact strand counted once. The mean over the complete bimodal network and its associated field range are
\begin{equation}
\langle\ell\rangle
=
w_1\langle\ell_1\rangle+w_2\langle\ell_2\rangle,
\qquad
\frac{R_0}{\xi}\simeq c_\ell\frac{\langle\ell\rangle}{\xi},
\label{eq:SI_mesh_range}
\end{equation}
where the single mesh size $\xi$ is the average intercrosslinker distance within the first coordination shell of the complete equilibrated reference network, and the same $c_\ell=1.54$ is used as for mono- and polydisperse networks. The weights in the mean strand end-to-end distance are the same strand-count fractions used in the reserve distribution. Substitution into Eq.~\eqref{eq:SI_bond_length_scaling} gives
\begin{equation}
\frac{\ell_{fc}}{\xi}
\sim
1+C^\star
\frac{w_1\langle\ell_1\rangle+w_2\langle\ell_2\rangle}{\xi}
\Delta_{\log}.
\label{eq:SI_bimodal_mesh_scaling}
\end{equation}
For two narrow populations, Eq.~\eqref{eq:SI_bimodal_narrow} makes the effects of mean strand end-to-end distance and prestretch contrast explicit,

\begin{figure}[t!]
    \centering
    \includegraphics[width=1\linewidth]{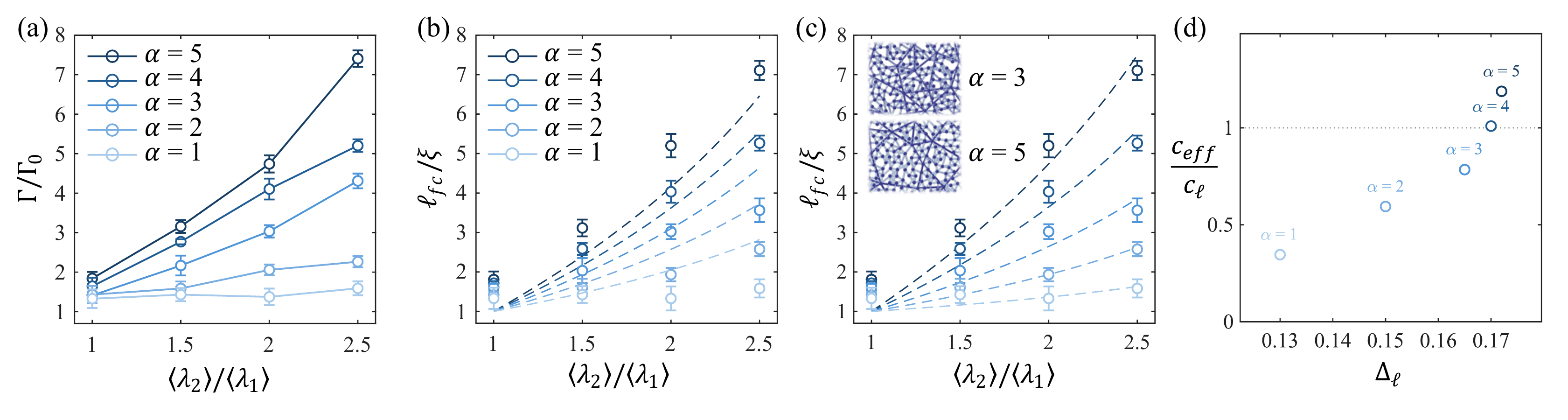}
    \caption{(a) Normalized toughness $\Gamma/\Gamma_0$ is plotted versus prestretch ratio $\langle\lambda_2\rangle/\langle\lambda_1\rangle$ for different values of strand length ratio $\alpha$. (b) Normalized fractocohesive length $\ell_{fc}/\xi$ is plotted versus prestretch ratio $\langle\lambda_2\rangle/\langle\lambda_1\rangle$ for different values of strand length ratio $\alpha$. The dashed scaling curves are plotted with fixed $c_\ell$. (c) Normalized fractocohesive length $\ell_{fc}/\xi$ is plotted versus prestretch ratio $\langle\lambda_2\rangle/\langle\lambda_1\rangle$ for different values of strand length ratio $\alpha$. The dashed scaling curves are plotted with effective coefficient $c_\ell=c_{eff}$ tuned for each value of $\alpha$. (d) The ratio of tuned constant to fixed constant $c_{eff}/c_\ell$ is plotted against the coefficient of variation of strand length distribution $\Delta_\ell$.}
    \label{fig:bimodal_scaling_with_fit_SI}
\end{figure}
\begin{equation}
\frac{\ell_{fc}}{\xi}
\sim
1+C^\star
\frac{w_1\langle\ell_1\rangle+w_2\langle\ell_2\rangle}{\xi}
\sqrt{w_1w_2}
\left|\ln\!\left(\frac{\bar\rho_1}{\bar\rho_2}\right)\right|.
\label{eq:SI_bimodal_bond_length_narrow}
\end{equation}
For finite population widths, Eq.~\eqref{eq:SI_bimodal_mesh_scaling} uses the full $\Delta_{\log}$ from Eq.~\eqref{eq:SI_total_log_variance}. Internal reserve variation can therefore contribute to damage broadening even when the mean prestretches coincide. For any number of strand populations, the same construction applies with $\langle\ell\rangle=\sum_s w_s\langle\ell_s\rangle$.\\

\noindent The ratio of mean strand end-to-end distances, $\alpha=\langle\ell_2\rangle/\langle\ell_1\rangle$, characterizes the geometric contrast between the two populations. Using $w_1+w_2=1$, the whole-network mean becomes $\langle\ell\rangle=\langle\ell_1\rangle[1+w_2(\alpha-1)]$, so Eq.~\eqref{eq:SI_mesh_range} gives $R_0/\xi\simeq c_\ell(\langle\ell_1\rangle/\xi)[1+w_2(\alpha-1)]$. Thus, the effect of $\alpha$ on the field range depends on both the relative strand abundance and the mean span of population 1 relative to the mesh size. The general estimate in Eq.~\eqref{eq:SI_bimodal_mesh_scaling} uses the population mean strand end-to-end distances directly. When comparing multiple realizations, the estimate is evaluated using the mean strand end-to-end distance and prestretch while assuming narrow reserve spread for each realization before averaging the predicted lengths.\\

\noindent Figure~\ref{fig:bimodal_scaling_with_fit_SI}(a) shows that the toughening associated with increasing prestretch contrast depends strongly on the strand-span contrast. When $\alpha=1$, increasing the mean prestretch ratio produces only a weak change in toughness, whereas the response becomes progressively stronger at larger $\alpha$. Thus, within the investigated family, prestretch contrast contributes more effectively to fracture resistance when accompanied by geometric contrast.\\

\noindent We first evaluate the fracture-length scaling using the fixed coefficients $c_\ell=1.54$ and $C=1.4007$ [Fig.~\ref{fig:bimodal_scaling_with_fit_SI}(b)]. The estimate captures the overall increase in fractocohesive length but overpredicts the response at larger prestretch contrasts for the low-$\alpha$ networks, particularly $\alpha=1$ and $2$. This discrepancy suggests that the mean strand span and logarithmic reserve spread do not fully describe the architecture dependence of their conversion into distributed damage. To account for this dependence, we relax the fixed-coefficient approximation for the bimodal fracture-length estimate by introducing $c_{\mathrm{eff}}$,
\begin{equation}
\frac{\ell_{fc}}{\xi}\sim1+2C c_{\mathrm{eff}}\frac{\langle\ell\rangle}{\xi}\Delta_{\log},\qquad C^\star=2C c_{\mathrm{eff}}. 
\label{eq:SI_bimodal_effective_scaling}
\end{equation}
A single $c_{\mathrm{eff}}$ is fitted for each $\alpha$ across the investigated prestretch ratios, while retaining the same width convention, strand statistics, and unit baseline. The resulting curves improve agreement with the simulation data [Fig.~\ref{fig:bimodal_scaling_with_fit_SI}(c)]. Here $c_{\mathrm{eff}}$ characterizes the effective conversion into fracture-active thickness and it is not an independent measurement of $R_0$.\\

\noindent The increase of $c_{\mathrm{eff}}/c_\ell$ with $\Delta_\ell$ [Fig.~\ref{fig:bimodal_scaling_with_fit_SI}(d)] is consistent with strand-span heterogeneity helping sustain heterogeneous loading during deformation and rupture. Such persistence could allow highly prestretched strands to undergo distributed scission while the less prestretched population maintains load-bearing connectivity. This stabilization mechanism provides a physical interpretation of the observed architecture dependence, although its direct verification would require tracking the population loading distributions during deformation.

\subsection{Extension to distributed rupture thresholds}

The same mechanism applies when variations in the rupture threshold produce the reserve distribution, with $\rho_{ij}=\lambda_{c,ij}-\lambda_{ij}>0$ measuring the remaining stretch of each strand. For a strength-based lattice model, the reserve and the crack-induced increment are instead expressed in stress units, so a weak ligament can be activated by a smaller stress increase farther from the crack.\\

\begin{figure}[t!]
\centering
\includegraphics[width=0.94\linewidth]{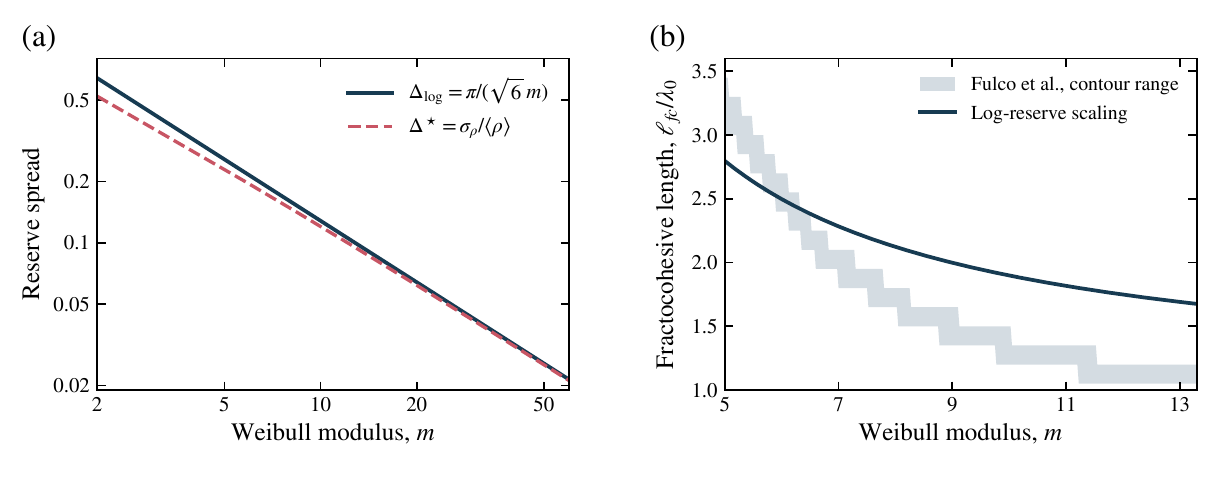}
\caption{\textbf{Extension to distributed rupture strength.} (a) The coefficient of variation $\Delta^\star$ and the exact logarithmic reserve width $\Delta_{\log}=\pi/(\sqrt{6}\,m)$ for Weibull strengths at zero reference stress. (b) Log-reserve scaling from Eq.~\eqref{eq:SI_weibull_length}, with fixed $C=1.4007$ and $R_0/\lambda_0=2.5$, compared with the normalized fractocohesive lengths in Fig.~4(b) of Fulco \textit{et al.}~\cite{fulco2025fracturedisorderedstochasticlattice}. The shaded band represents the range read from their contours across geometric disorder, including finite color resolution, rather than statistical uncertainty. All lengths in panel (b) use their ordered, deterministic reference $\lambda_0$.}
\label{fig:strength_scaling}
\end{figure}

\noindent Fulco \textit{et al.}~\cite{fulco2025fracturedisorderedstochasticlattice} consider Weibull-distributed ligament strengths $S$, with strength scale $S_0$ and modulus $m$, where decreasing $m$ broadens the strength distribution. At zero reference stress the reserve is $S$, and its logarithmic width follows directly from the Weibull distribution,
\begin{equation}
\boxed{\Delta_{\log}=\operatorname{std}\!\left[\ln(S/S_0)\right]=\frac{\pi}{\sqrt{6}\,m}.}
\label{eq:SI_weibull_log_width}
\end{equation}
The strength scale cancels from this spread because it measures relative contrasts among rupture reserves, while normalization by $S_0$ makes the logarithm dimensionless. Equation~\eqref{eq:SI_weibull_log_width} is exact for all $m>0$, whereas the coefficient of variation $\Delta^\star=\sigma_S/\langle S\rangle$ approaches the same expression only for narrow strength distributions at large $m$, as shown in Fig.~\ref{fig:strength_scaling}(a).\\

\noindent To compare with their reported fracture lengths, we use the ordered, deterministic reference length $\lambda_0$ as the localized baseline in place of the polymer mesh scale in Eq.~\eqref{eq:SI_unified_scaling}, giving
\begin{equation}
\frac{\ell_{fc}^{\mathrm{pred}}}{\lambda_0}\sim1+2C\frac{R_0}{\lambda_0}\frac{\pi}{\sqrt{6}\,m}.
\label{eq:SI_weibull_length}
\end{equation}
Figure~\ref{fig:strength_scaling}(b) compares this estimate with the reported normalized lengths using the same fixed $C=1.4007$ and an illustrative constant $R_0/\lambda_0=2.5$. Because their crack-tip deformation is described by a $K$-field and no exponential range is independently measured, this comparison illustrates the increase in fracture length with strength disorder without establishing quantitative agreement. Here $R_0/\lambda_0$ remains an illustrative effective-range parameter. The bond-length relation with $c_\ell=1.54$ is used for the present polymer networks and is not imposed on this lattice comparison.

\begin{figure*}[b!]
    \centering
    \includegraphics[width=1\linewidth]{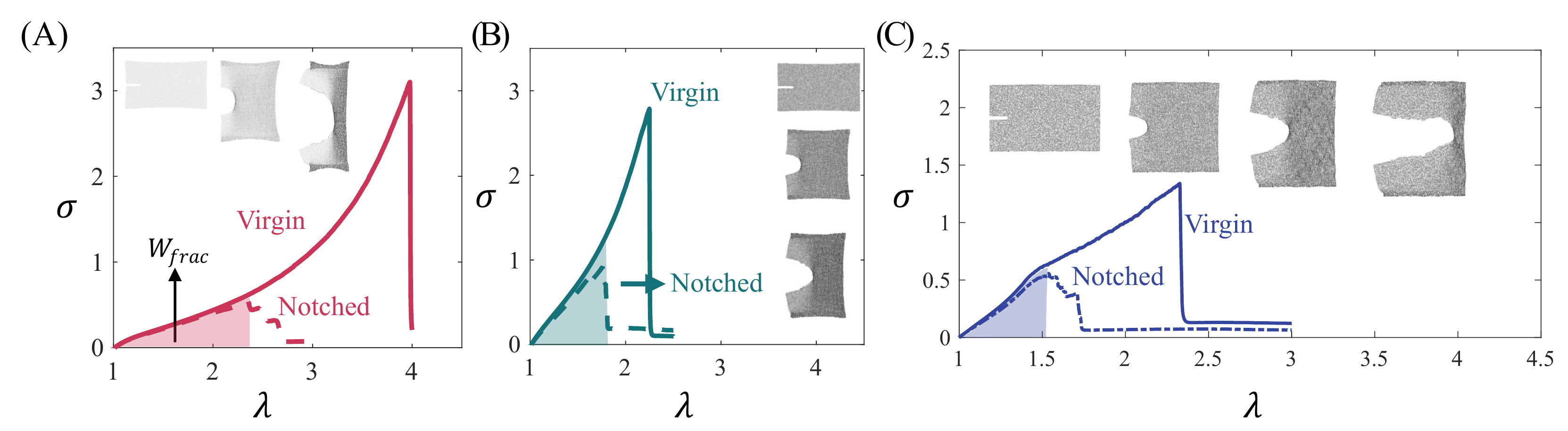}
    \caption{\footnotesize
    \textbf{Stress--stretch response of virgin and notched networks for different architectures.}
    (A) Monodisperse, (B) polydisperse, and (C) bimodal networks. Solid curves show the unnotched (virgin) response and dashed curves show the response of the corresponding notched samples. The shaded area indicates the work of fracture \(W_{\mathrm{frac}}\), defined as the area under the virgin stress--stretch curve up to the failure stretch of the corresponding notched sample. Insets show representative configurations of notched samples during deformation and crack propagation.
    }

    \label{fig:endmatter}
\end{figure*}

\section{Stress--stretch response of different architectures}

Figure~\ref{fig:endmatter} compares the stress--stretch response of virgin and notched samples for monodisperse, polydisperse, and bimodal network architectures. In each case, the solid curve denotes the response of the unnotched sample and the dashed curve denotes the response of the corresponding notched sample. The area under the virgin curve up to the failure stretch of the notched sample defines the work of fracture, \(W_{\mathrm{frac}}\), used in the main text to quantify the sensitivity of each architecture to pre-existing flaws.

The three architectures exhibit qualitatively distinct failure responses. In the monodisperse network, the virgin sample reaches the largest stretch and stress under homogeneous loading, while the corresponding notched sample fails at substantially lower stretch, consistent with strong flaw sensitivity. In the polydisperse network, both the virgin and notched responses occur at lower stress and stretch, but the difference between them is reduced over a broader range of deformation, consistent with the increased flaw tolerance discussed in the main text. The bimodal network shows the strongest reduction in virgin work density together with a pronounced loss of load-bearing capacity in the notched state, reflecting early activation of highly prestretched strands and subsequent crack growth.

The insets in Fig.~\ref{fig:endmatter} show representative configurations of the notched samples during deformation and fracture. These snapshots illustrate the distinct crack-growth morphologies associated with the three architectures and provide a visual complement to the corresponding stress--stretch curves.

\section{Stretch fields in front of crack}
\begin{figure*}[h!]
    \centering
    \includegraphics[width=0.7\linewidth]{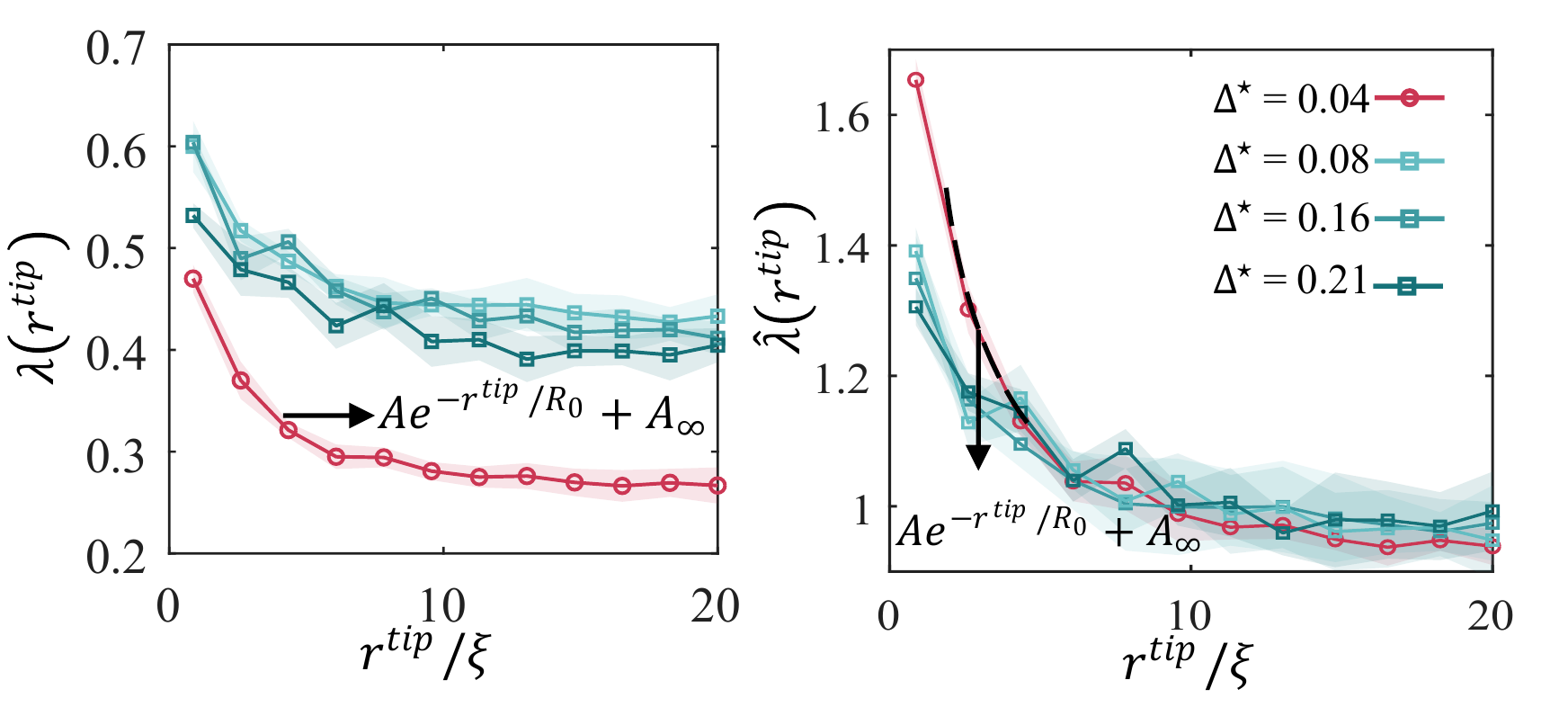}
    \caption{\footnotesize average bond stretch versus crack tip distance plotted in both original and normalized forms.}
    \label{fig:Stretch_field_SI}
\end{figure*}

In figure~\ref{fig:Stretch_field_SI}, we illustrate the same stretch maps showed in manuscript in both original and normalized values as a function of crack tip distance. It can be seen that increasing the prestretch heterogeneity ($\Delta^\star$), the stretch concentration near tip becomes smaller and also the decay lengthscale $R_0$ increases,

\bibliography{references.bib}